\documentclass[journal]{IEEEtran}
\usepackage{framed,multirow}
\usepackage{amssymb}
\usepackage{latexsym}
\usepackage{url}
\usepackage{xcolor}
\usepackage{hyperref}
\usepackage{amsmath,amssymb,amsfonts}
\usepackage{algorithmic}
\usepackage{algorithm}
\usepackage{graphicx}
\usepackage{color}
\usepackage{multirow}
\usepackage{longtable}
\usepackage{booktabs}
\usepackage{ragged2e}
\usepackage{pifont}
\usepackage{textcomp}
\usepackage[justification=centering]{caption}
\usepackage{cite}
\usepackage[figuresright]{rotating}
\usepackage{colortbl}
\definecolor{mygray}{gray}{.9}
\ifCLASSINFOpdf
\else
\fi
\begin{document}
%
\title{Towards Reliable Medical Image Segmentation by Modeling Evidential Calibrated Uncertainty}
%
%
%

\author{Ke Zou, Yidi Chen, Ling Huang, Nan Zhou, Xuedong Yuan, Xiaojing Shen~\IEEEmembership{Member, IEEE},\\ Meng Wang, Rick Siow Mong Goh, Yong Liu, Yih Chung Tham, Huazhu Fu~\IEEEmembership{Senior Member, IEEE}
\thanks{Manuscript received xxx, 2023; revised xxx, 2024. This work was supported by the H. Fu’s Agency for Science, Technology and Research (A*STAR) Central Research Fund (“Robust and Trustworthy AI system for Multi-modality Healthcare”), the National Research Foundation, Singapore under its AI Singapore Programme (AISG Award No: AISG2-TC-2021-003), the National Key Researchand Development Program of China under Grant 2020YFA0714003, and the China Scholarship Council (No. 202206240082).}
\thanks{K. Zou, N. Zhou and X. Yuan are with the National Key Laboratory of Fundamental Science on Synthetic Vision, the College of Computer Science, Sichuan University, Chengdu 610065, China. }
\thanks{Y. Chen is with the Department of Radiology, West China Hospital, Sichuan University, Chengdu 610065, China. }
\thanks{L. Huang is with the Saw Swee Hock School of Public Health, National University of Singapore, Singapore 117549. }
\thanks{X. Shen is with the Department of Mathematics, Sichuan University, Chengdu 610065, China. }
\thanks{M. Wang, R. Goh, Y. Liu, and H. Fu are with the Institute of High Performance Computing (IHPC), Agency for Science, Technology and Research (A*STAR), Singapore 138632. }
\thanks{Y. Chung is with the Centre for Innovation and Precision Eye Health, Yong Loo Lin School of Medicine, National University of Singapore, Singapore 117549 and the Singapore Eye Research Institute, Singapore National Eye Centre, Singapore 168751.}
\thanks{Xuedong Yuan and Huazhu Fu are the co-corresponding authors (e-mail: yxdongdong@163.com, hzfu@ieee.org). }}

\markboth{Journal of \LaTeX\ Class Files,~Vol.~14, No.~8, August~2015}%
{Shell \MakeLowercase{\textit{et al.}}: Bare Demo of IEEEtran.cls for IEEE Journals}
%



\maketitle

\begin{abstract}
Medical image segmentation is critical for disease diagnosis and treatment assessment. However, concerns regarding the reliability of segmentation regions persist among clinicians, mainly attributed to the absence of confidence assessment, robustness, and calibration to accuracy. To address this, we introduce DEviS, an easily implementable foundational model that seamlessly integrates into various medical image segmentation networks. DEviS not only enhances the calibration and robustness of baseline segmentation accuracy but also provides high-efficiency uncertainty estimation for reliable predictions. By leveraging subjective logic theory, we explicitly model probability and uncertainty for medical image segmentation. Here, the Dirichlet distribution parameterizes the distribution of probabilities for different classes of the segmentation results. To generate calibrated predictions and uncertainty, we develop a trainable calibrated uncertainty penalty. Furthermore, DEviS incorporates an uncertainty-aware filtering module, which designs the metric of uncertainty-calibrated error to filter out-of-distribution data. We conducted validation studies on publicly available datasets, including ISIC2018, KiTS2021, LiTS2017, and BraTS2019, to assess the accuracy and robustness of different backbone segmentation models enhanced by DEviS, as well as the efficiency and reliability of uncertainty estimation. Additionally, two potential clinical trials were conducted using the uncertainty-aware filtering module. The clinical application conducted on the Johns Hopkins OCT and Duke OCT-DME datasets demonstrated the effectiveness of the model in filtering out-of-distribution data. The second trial evaluated its efficacy in filtering high-quality data on the FIVES datasets. At last, the proposed DEviS method was extended to semi-supervised medical image segmentation, where it exhibited strong robustness under noisy conditions. Our code has been released in https://github.com/Cocofeat/DEviS.
\end{abstract}


\IEEEpeerreviewmaketitle

\section{Introduction}
\IEEEPARstart{M}{edical} image segmentation, benefit from deep learning techniques, has ushered in a paradigm shift in quantitative pathological assessments~\cite{deng2024cross}, diagnostic support systems~\cite{zhang2023trustworthy}, and tumor analysis~\cite{bilic2023liver}. In response to identifying reliable medical image segmentation regions, the focus has transcended mere accuracy. A reliable medical image segmentation model assumes a pivotal role in establishing a foundation of trust and confidence between healthcare professionals and patients. It provides efficient pixel-level confidence among healthcare professionals while delivering robust results. In addition to strive for a closer alignment of network outputs with the ground truth, a reliable model also focuses on delivering accurate and well-calibrated predictions. This allows the model to provide indications of when its output can be trusted, effectively avoiding situations where the model produces overconfident predictions~\cite{2017ece}. Significantly, the reliable model provides timely warnings when confronted with data  high uncertainty, thereby recognizing the potential limitations and reliability of its predictions. With this paper, we embark on a journey to develop a reliable foundational model for medical image segmentation, facilitating robust segmentation and confident region identification. By employing calibrated uncertainty estimation and validating partial clinical applications, our ultimate aim is to foster trust between healthcare professionals and deep learning models.

Remarkable advancements have been achieved in the endeavor to enhance the accuracy of deep network architectures for medical image segmentation. Fully Convolutional Networks (FCN) have been developed to achieve end-to-end semantic segmentation with impressive results~\cite{FCN15}. Based on FCN, the U-Net~\cite{Unet15} model and its variants~\cite{AttentionU18, VNet} were proposed to improve feature representations and segmentation outcomes. Meanwhile, highly expressive transformers~\cite{transformer17,zhou2023nnformer} have gained success in computer vision and medical image segmentation. More recently, Models such as MedSAM~\cite{MedSAM} and Vision Mamba~\cite{zhang2024vm,xing2024segmamba} have further advanced segmentation performance by leveraging specialized architectures and adaptive mechanisms. While existing research primarily focuses on improving segmentation accuracy, two critical limitations hinder further development. First, current backbone models largely overlook ambiguous decisions that AI systems may produce, lacking principled methods to quantify uncertainty needed for safe clinical application. Second, fundamental segmentation frameworks struggle to handle noise perturbations and identify high-uncertainty samples, which are essential for robust real-world deployment.


Uncertainty estimation methods play a pivotal role in assessing model confidence. According to~\cite{zou2023review}, uncertainty quantification in medical domain includes Bayesian~\cite{2018probabilisticU, roy2019bqnat} and non-Bayesian~\cite{wang2019aleatoric, ensemble17, evidential18,ICMLdeterministic20} based methods. Bayesian-based methods~\cite{2018probabilisticU} enables the segmentation networks to learn a distribution over the model parameters with uncertainty, which are computationally expensive~\cite{16dropout}. Monte Carlo Dropout-based methods~\cite{MIA2020exploringDropSeg} alleviate this problem by introducing dropout to approximate uncertainty from multiple predictions. Non-Bayesian based methods have been developed, which mainly include test time augmentation (TTA)~\cite{wang2019aleatoric}, deep ensemble (DE)~\cite{ensemble17}, deterministic~\cite{ICMLdeterministic20}, and evidential~\cite{evidential18} based methods. In this paper, we focus on evidence-based deep learning methods~\cite{evidential18} applied to medical image segmentation, aiming to provide accurate and reliable segmentation. Traditional medical image segmentation only offers approximate segmentation results without providing confidence measures for each pixel. Consequently, when encountering anomalous samples such as noise, its segmentation performance deteriorates, failing to indicate which pixels are reliable. Ensemble-based medical image segmentation method trained multiple models to produce multiple predictions for obtaining average prediction results and uncertainty. However, this method incurs high training costs for uncertainty estimation and yields poorer results under noisy samples. In contrast, the reliable medical image segmentation method proposed in this paper, not only provides robust segmentation results but also offers uncertainty estimations for each pixel. Moreover, this uncertainty estimation only requires one forward pass, significantly enhancing computational efficiency. Additionally, previous research primarily focused on utilizing uncertainty to improve segmentation performance, overlooking the crucial aspect of model calibration. Lastly, unlike the aforementioned uncertainty estimation methods, which are less utilized in healthcare for clinical applications, this paper extends the proposed method to filtering out-distribution data.

In this study, we present a \textbf{D}eep \textbf{Evi}dential \textbf{S}egmentation model (\textbf{DEviS}) for identifying reliable medical image segmentation regions that seamlessly integrates with existing base frameworks, offering efficient confidence assessment, robustness, and well-calibrated results. Our work also explores the potential of DEviS in clinical applications. The key contributions of this study are fourfold: 
\begin{itemize}
    \item [1)] We derive probabilities and uncertainties for different class segmentation problems via Subjective Logic (SL) theory, where the Dirichlet distribution parameterizes the distribution of probabilities for different classes of the segmentation results.
    \item [2)] A trainable calibrated uncertainty penalty (CUP) is developed to generate more calibrated confidence and maintain the segmentation performance of the base network.
    \item [3)] An uncertainty-aware filtering (UAF) strategy is proposed to facilitate the translation of DEviS into clinical applications. 
    \item [4)] We conducted extensive experiments on four publicly available datasets to evaluate the model with different backbone's accuracy and robustness in predicting both standard and adversarial samples, including those with noise, blur, and random masking.  Additionally, we assessed the efficiency and reliability of uncertainty estimation.
    \item [5)] Four other publicly available datasets were used to simulate clinical applications, validating the effectiveness of the UAF strategy through image quality assessment and out-of-distribution (OOD) detection experiments, as well as evaluating the generalization capabilities of the CUP strategy. Furthermore, the proposed DEviS method was extended to semi-supervised medical image segmentation, where it demonstrated strong robustness under noisy conditions, further supporting its practical applicability.
\end{itemize}
To reiterate, we establish an easily portable framework for achieving reliable medical image segmentation. Compared to our previous conference version~\cite{zou2022tbrats}, we have made substantial improvements in several key aspects. First, we introduce a trainable calibrated uncertainty penalty to produce more reliable confidence estimates, thereby preserving the segmentation performance of the base network. Second, we design an UAF strategy to facilitate the integration of DEviS into clinical workflows. We also conduct clinical studies involving OOD data detection and image quality assessment, viewed through the lens of segmentation uncertainty. Finally, we perform extensive experiments on diverse datasets to comprehensively evaluate the accuracy, robustness, and reliability of our model.

We further compare our work with several representative medical image segmentation methods recently published in this journal. GT-DLA-dsHFF~\cite{li2022global} enhances fine-grained vessel segmentation by integrating global transformers, dual local attention, and hierarchical feature fusion. In contrast, our model generalizes across medical domains with a reliable and easily integrable foundational design. Pang et al.\cite{pang2022beyond} proposed GER-UNet, which encodes geometric symmetries to improve segmentation accuracy. Instead, we focus on segmentation reliability through uncertainty modeling and calibrated prediction. Huang et al.\cite{huang20203D} introduced 3D RU-Net for efficient colorectal tumor segmentation from 3D MR images. Our approach, however, supports a broad range of architectures and leverages uncertainty estimation to highlight ambiguous tumor boundaries often missed by conventional segmentation models.

\section{Proposed method}
In our proposed DEviS model, we employ an arbitrary network structure as the backbone network for extracting deep evidential features, which can be U-Net~\cite{Unet15,isensee2021nnu}, its variants~\cite{VNet, AttentionU18}, or transformer-based~\cite{zhou2023nnformer} and vision mamba-based~\cite{zhang2024vm} methods. To address the issue of over-confidence, DEviS utilizes the softplus activation function layer instead of the softmax layer. Additionally, it utilizes the Dirichlet distribution to generate a predictive distribution rather than a single output, enhancing the reliability of the segmentation results~\cite{evidential18}. Furthermore, SL~\cite{SL16} is incorporated to establish probability and uncertianty for different segmentation classes, enabling evidential uncertainty estimation. To enhance the reliability of model predictions in medical diagnosis, a specifically designed CUP for well-calibrated medical image segmentation is developed. Moreover, a UAF strategy is devised to facilitate subsequent clinical tasks by detecting and indicating medical data with its uncertainty. Ultimately, a comprehensive training loss function is formulated, encompassing medical image segmentation, uncertainty estimation, and uncertainty calibration, to ensure reliable medical image segmentation. The following section presents preliminary knowledge on DEviS.

\begin{figure*}[!t]
\centering
\includegraphics[width=1\linewidth]{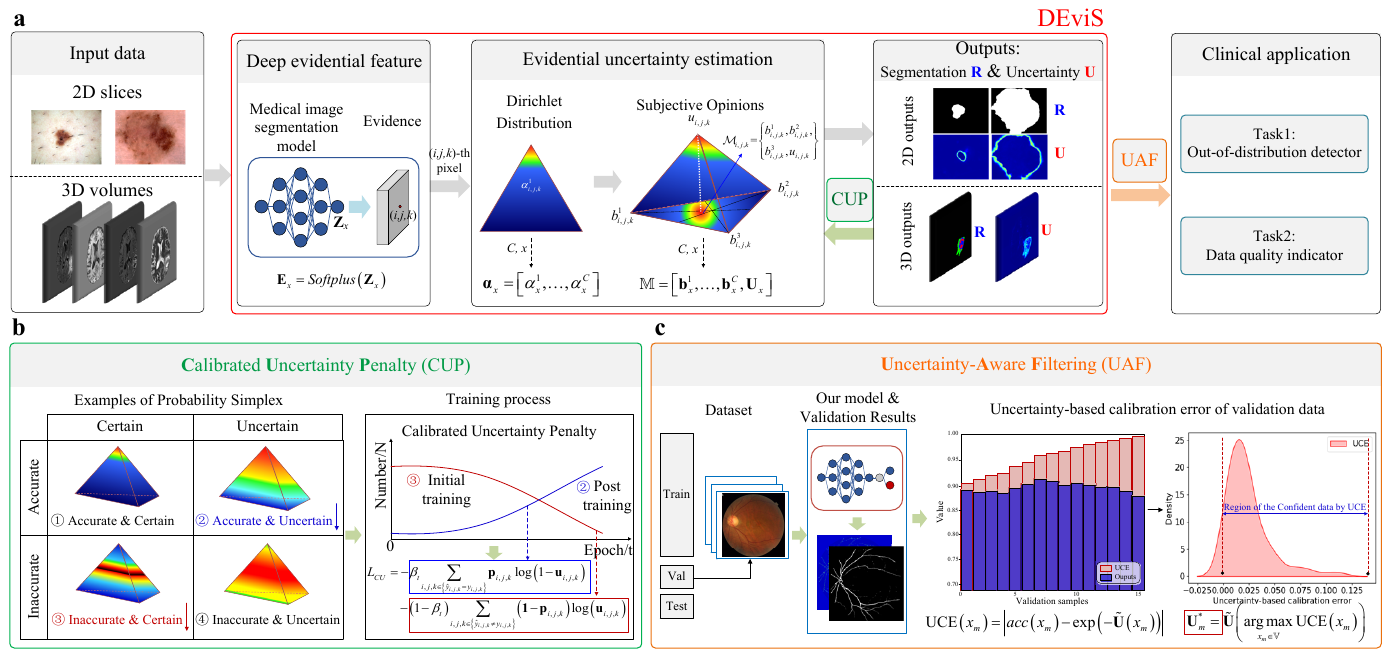}
\caption{\textbf{a}. The overall framework of DEviS. \textbf{b}. \textcolor{blue}{R} and \textcolor{red}{U} denote the prediction and uncertainty map. The training process of calibrated uncertainty penalty. \textbf{c}. Schematic for uncertainty-aware filtering.}
\label{F_1}
\end{figure*}
\subsection{Preliminaries: Multinomial opinions and Dirichlet distribution}
In multi-class classification problems, the softmax function is widely used in deep neural networks to convert raw outputs into probability distributions over mutually exclusive classes. Mathematically, these probabilities can be interpreted as parameters of a multinomial distribution, representing the likelihood of different outcomes~\cite{SL16}. Building on this, J{\o}sang et al. characterized multi-class probabilities using multinomial opinions, proposing that probabilities consist of both belief mass and uncertainty mass. Accordingly, For $\forall x \in \mathbb{X}$, the projected probability distribution of multinomial opinions is defined by:
\begin{equation}
\label{E_2}
\boldsymbol{p}_x=\boldsymbol{b}_x+\boldsymbol{r}_x \boldsymbol{U}_x,
\end{equation}
where $\boldsymbol{b}_x$, $\boldsymbol{r}_x$ and $\boldsymbol{U}_x$ are the belief mass distribution, base rate distribution and the uncertainty mass distribution over $x$, respectively. Here, the variable $x$ in this paper represents any dimensionality medical image, including two-dimensional, three-dimensional, or even multi-dimensional images.

While softmax is effective, extensive researches have highlighted its tendency to produce overconfident predictions~\cite{2017ece,ICMLdeterministic20}. To address this issue, we model network predictions using a Dirichlet distribution instead of relying solely on individual class probabilities. As the conjugate prior of the multinomial distribution, the Dirichlet distribution allows for a more principled way to quantify uncertainty, thereby mitigating the overconfidence problem~\cite{2017ece,evidential18,2020trusted}. The definition of Dirichlet distribution is as follows.

\textbf{Definition 2.1 (Dirichlet Distribution)} The Dirichlet PDF $D({{\boldsymbol{p}}_x}\mid {{\boldsymbol{\alpha }}_x})$ can be used to represent probability density over ${{\boldsymbol{p}}_x}$, which is given by:
\begin{equation}
\label{E_3}
D({{\boldsymbol{p}}_x}\mid {{\boldsymbol{\alpha }}_x}) = \left\{ {\begin{array}{*{20}{l}}
{\frac{1}{{B({{\boldsymbol{\alpha }}_{x}})}}\prod\limits_{c = 1}^C {{{\left( {p_x^c} \right)}^{\alpha _x^c - 1}}} }&{{\rm{for}}\quad {{\boldsymbol{p}}_x} \in {{S}_x^C}}\\
0&{{\rm{otherwise}}}
\end{array}} \right.,
\end{equation}
where Dirichlet distribution with parameters ${{\boldsymbol{\alpha }}_x} = \left[ {\alpha _x^1, \ldots ,\alpha _x^C} \right]$ is considered as basic belief assignment. $B({\boldsymbol{\alpha }_x})$ is the $C$-dimensional multinomial beta function, and ${{S}_x^C}$ is the C-dimensional unit simplex, given by:
\begin{equation}
\label{E_4}
{{S}_x^C} = \left\{ {{\boldsymbol{p}_x}\left| {\sum\limits_{c = 1}^C {p_x^c}  = 1\quad {\rm{and}}\quad 0 \le p_x^1, \ldots ,p_x^C \le 1} \right.} \right\}.
\end{equation}
Following \cite{kruschke2010bayesian,SL16}, a multinomial opinion corresponds to a Dirichlet probability density function (PDF) over \(X\), allowing for a bijective mapping between Dirichlet PDFs and multinomial opinions. Then, the total strength ${\boldsymbol{\alpha }}_x$ can be denoted as:
\begin{equation}
\label{E_5}
{{\boldsymbol{\alpha }}_x} = {{\bf{E}}_x} + {W}{{\bf{r}}_x},
\end{equation}
As noted by~\cite{SL16}, a larger selection of W reduces the impact of new observational evidence on the Dirichlet PDF and the estimated probability distribution. Following the approach in~\cite{evidential18, 2020trusted},  the base rate distribution is set to a uniform value of 1. Accordingly, the formula simplifies to ${{\boldsymbol{\alpha }}_x} = {{\bf{E}}_x} + {{\boldsymbol{1}}}$ in this paper. This simplification is grounded in the properties of the Dirichlet distribution, where ${{\boldsymbol{1}}}$ represents a uniform prior. Furthermore, when the evidence is zero, the Dirichlet distribution naturally reduces to this form.

\subsection{Constructing DEviS \label{M_2}} 
\textbf{1) Deep evidential feature extraction.} U-Net~\cite{Unet15,isensee2021nnu}, its variants~\cite{VNet, AttentionU18}, transformer~\cite{zhou2023nnformer} and Vision mamba~\cite{zhang2024vm} have seen recent widespread used across medical image modalities. We thus employed one of them as our backbone for capturing contextual information. For U-Net based methods, the backbones only performed down-sampling three times to achieve a balance between GPU memory usage and segmentation accuracy. For the transformer-based method, we follow its design~\cite{zhou2023nnformer}. DEviS can freely choose any backbone to extract image features. We only use its decoder output feature vector without the softmax layer. For a random image $x$ in a medical image domain $\mathbb{X}$, this process can be defined as:
\begin{equation}
\label{E_0}
{{\bf{Z}}_x} = {f}\left( x \right)
\end{equation}
Where ${f}\left(  \cdot  \right)$ is any network backbone without the softmax layer. In this study, we assessed the performance of the different general backbones (U-Net~\cite{Unet15}, V-Net~\cite{VNet}, Attention-UNet \cite{AttentionU18}, nnU-Net~\cite{isensee2021nnu}), nnformer~\cite{zhou2023nnformer} and Vision Mamba U-Net~\cite{zhang2024vm}. For typical medical image segmentation tasks \cite{2020CA,HDenseU,TransBTS}, the predictions are usually carried by the softmax layer as the final layer. As mentioned in~\cite{ICMLdeterministic20}, the softmax layer has a tendency to display high confidence even for wrong predictions. DEviS mitigates this issue by utilizing the softplus activation function layer instead of the softmax layer. Therefore, following the output of any neural network, a softplus activation function layer ($Softplus$) is applied to ensure that the network output is non-negative, which is considered as evidence for the backbone feature output.
\begin{equation}
\label{E_1}
{{\bf{E}}_x} = Softplus\left( {{{\bf{Z}}_x}} \right)
\end{equation}
\textbf{2) Evidential uncertainty estimation.} One of the generalizations of Bayesian theory for subjective probability is the Dempster-Shafer Theory (DST) \cite{DS08}. The Dirichlet distribution is formalized as the belief distribution of DST over the discriminative framework in the SL~\cite{SL16}. For medical image segmentation, we define the DEviS framework through SL \cite{SL16}, which derives the probability and uncertainty of different class segmentation problem based on deep evidential features. In Fig. 1(a), we illustrate the uncertainty estimation process 3D $C$-class medical image segmentation task. Given a 3D image input $x$ and evidential feature output ${{\bf{E}}_x}{\rm{ = }}\left\{ {{\bf{E}}_x^1, \ldots ,{\bf{E}}_x^c} \right\}$,  where ${{\bf{E}}_x^c}$, $c = 1,2 \ldots ,C$ denotes the evidence for $c$-th class. The SL provides a belief mass ${{\bf{b}}_x} = \left\{ {{\bf{b}}_x^c} \right\}_{c = 1}^C$ and an uncertainty mass ${{\bf{u}}_x}$ for different classes of segmentation results. The $C + 1$ mass values are all non-negative and their sum is one. Specifically, for the $(i,j,k)$-th pixel, it has the following definition:
\begin{equation}
\label{E_6}
{u_{i,j,k}} + \sum\limits_{c = 1}^C {b_{i,j,k}^c}  = 1,
\end{equation}
where $b_{_{i,j,k}}^c \ge 0 $ and ${u_{i,j,k}} \ge 0$ denote the probability of the $(i,j,k)$-th pixel for the $c$-th class and the overall uncertainty of the $(i,j,k)$-th pixel in $x$, respectively, where ${u_{i,j,k}} \in {\bf{u}}_x$ and $b_{_{i,j,k}}^c \in {{\bf{b}}_x^c}$. SL then associates the evidence ${{e_{i,j,k}^c}}$ having the Dirichlet distribution with the parameters ${\alpha _{i,j,k}^c}={{e_{i,j,k}^c}}+1$, where ${{e_{i,j,k}^c}} \ge 0 $ and ${{e_{i,j,k}^c}}\in {{\bf{E}}_x^c}$. After that, the belief mass and the uncertainty of the $(i,j,k)$-th pixel can be denoted as:
\begin{equation}
\label{E_7}
b_{i,j,k}^c = \frac{{e_{i,j,k}^c}}{S} = \frac{{\alpha _{i,j,k}^c - 1}}{S} \quad{\rm{and }}\quad{u_{i,j,k}} = \frac{C}{S},
\end{equation}
where $S = \sum\limits_{c = 1}^C {\alpha _{i,j,k}^c}  = \sum\limits_{c = 1}^C {\left( {e_{i,j,k}^c + 1} \right)} $ denotes the Dirichlet strength. The more evidence of the $c$-th class obtained by the $(i,j,k)$-th pixel, the greater its probability. Vice versa. 

For clarity, we provide a example that demonstrates the formulation in the context of a three-category 3D medical image segmentation task. Let us consider the evidence output ${{\bf{e}}_{i,j,k}} = \left[ {40,1,1} \right]$ for the $(i,j,k)$-th pixel. The corresponding Dirichlet distribution parameter is ${{\bf{\alpha }}_{i,j,k}} =  \left[ {41,2,2} \right]  $. In accordance with the SL theory~\cite{SL16}, the three categories of subjective opinions are represented by ${{{b}}_{i,j,k}^1} =  \left[40,0,0\right]$, ${{{b}}_{i,j,k}^2} =  \left[ 0,1,0 \right]$, and ${{{b}}_{i,j,k}^3} =  \left[0,0,1\right]$, respectively. Additionally, the uncertainty mass is given by ${{u}_{i,j,k}} =  \left[0.067\right]$. Hence, the point ${\cal M}$ on the simplex represents the opinions of the $(i,j,k)$-th pixel as ${\mathcal M}{\rm{ = }}\left\{ {{{b}}_{i,j,k}^1,{{b}}_{i,j,k}^2,{{b}}_{i,j,k}^3,{u_{i,j,k}}} \right\}$. This observation indicates that there is a significant amount of evidence supporting the classification of the $(i, j, k)$-th pixel as the first class, with an associated uncertainty value of 0.067 for the pixel. This formulation can be extended to all pixels in $x$, allowing the estimation of the confidence for each individual pixel. The subjective opinion of image $x$ is then obtained as $ \mathbb{M} = \left[ {{\bf{b}}_x^1, \ldots ,{\bf{b}}_x^C,{{\bf{u}}_x}} \right]$.

\textbf{3) Calibrated uncertainty penalty.\label{M_3}} 
Though DEviS provides a means for directly learning evidential uncertainty without sampling, the uncertainty might not be totally calibrated to enhance the reliability of model predictions. As noted in~\cite{AvUeval18, AvU20}, a reliable and well-calibrated model should exhibit high uncertainty when making inaccurate predictions and high confidence when making accurate ones. Therefore, we first introduce the CUP score to optimize the model for well-calibrated uncertainty in the medical domain, as defined by the following equation: $\rm{CUP} = \frac{N_{AC} + N_{IU}}{N_{AC} + N_{AU} + N_{IC} + N_{IU}}$ where ${N_{AC}}$, ${N_{AU}}$, ${N_{IC}}$ and ${N_{IU}}$ denote the number of samples for the network output category of Accurate and Certain (A\&C), Accurate and Uncertain (A\&U), the Inaccurate and Certain (I\&C) and the Inaccurate and Uncertain (I\&U), respectively (Examples of probability simplex can be seen in Fig.~\ref{F_1} \textbf{b} CUP). Fig.~\ref{F_1} \textbf{b} illustrates a toy example of these four possible model outputs. To calibrate uncertainty in predictions, the model is encouraged to learn a tilted and sharp Dirichlet simplex for accurate predictions (Fig.~\ref{F_1} \textbf{b} \ding{172} A\&C). Conversely, for incorrect predictions, the model is expected to produce an unbiased and flat Dirichlet simplex (Fig.~\ref{F_1} \textbf{b} \ding{175} I\&U). Similarly, for clarity, we present four examples corresponding to Fig.~\ref{F_1} \textbf{b}, illustrating the probability simplex in three types of 3D medical image segmentation tasks. Specifically, for A\&U, we assume that the obtained ${e_{i,j,k}} =  \left[ {7,1,1} \right]$, following equations~\ref{E_6} and~\ref{E_7} corresponds to the Dirichlet parameter ${{\alpha }_{i,j,k}} =  \left[ {8,2,2} \right]$, with a high uncertainty of ${{u}_{i,j,k}} =  \left[0.25\right]$. For I\&C, given ${e_{i,j,k}} =  \left[ {40,40,40} \right]$ and its corresponding Dirichlet parameter ${{\alpha }_{i,j,k}} =  \left[ {41,41,41} \right]$, but with a low uncertainty ${{u}_{i,j,k}} =  \left[0.024\right]$. To achieve this, we propose calibrating the model’s training by minimizing the expected values of A\&U and I\&C cases (Fig.~\ref{F_1} \textbf{b} \ding{173} and \ding{174}), thereby promoting the desired behavior in the other two cases. We define proxy functions to approximate them as given by:
\begin{equation}
\label{E_8}
\begin{array}{l}
{N_{AC}} = \sum\limits_{i,j,k \in \left\{ {{{\hat y}_{i,j,k}} = {y_{i,j,k}}} \right\}} {\sum\limits_{c = 1}^C {b_{i,j,k}^c\log \left( {{u_{i,j,k}}} \right)} } ;\\
{N_{AU}} = \sum\limits_{i,j,k \in \left\{ {{{\hat y}_{i,j,k}} = {y_{i,j,k}}} \right\}} {\sum\limits_{c = 1}^C {b_{i,j,k}^c\log \left( {1 - {u_{i,j,k}}} \right)} } ;\\
{N_{IC}} = \sum\limits_{i,j,k \in \left\{ {{{\hat y}_{i,j,k}} \ne {y_{i,j,k}}} \right\}} {\sum\limits_{c = 1}^C {\left( {1 - b_{i,j,k}^c} \right)\log \left( {{u_{i,j,k}}} \right)} } ;\\
{N_{IU}} = \sum\limits_{i,j,k \in \left\{ {{{\hat y}_{i,j,k}} \ne {y_{i,j,k}}} \right\}} {\sum\limits_{c = 1}^C {\left( {1 - b_{i,j,k}^c} \right)\log \left( {1 - {u_{i,j,k}}} \right)}},
\end{array}
\end{equation}

Based on the above formulation, ideally, we expect the model to produce low uncertainty for accurate predictions and high uncertainty for inaccurate predictions. Hence, a reliable and well-calibrated model will yield a higher CUP score.  Furthermore, this paper proposes a differentiable approximation of the CUP score and introduces a trainable uncertainty calibration loss $ \mathcal{L}_{CUP} $ by leveraging the negative log-CUP score, as shown in the following equation:
\begin{equation}
\label{E_9}
\begin{array}{l}
{{\mathcal L}_{CUP}} =  - \log \left( {\frac{{{N_{AC}} + {N_{IU}}}}{{{N_{AC}} + {N_{AU}} + {N_{IC}} + {N_{IU}}}}} \right)\\
{\rm{\quad \quad \quad = }}\log \left( {1 + \frac{{{N_{AU}} + {N_{IC}}}}{{{N_{AC}} + {N_{IU}}}}} \right)
\end{array}
\end{equation}
This loss term relies on a theoretically sound loss-calibrated approximate inference framework \cite{2011Approximate} as a utility-dependent penalty term to achieve well-calibrated uncertainty in the medical domain. The objective is to maintain low uncertainty for accurate predictions and high uncertainty for inaccurate predictions during training, thereby ensuring well-calibrated uncertainty. This not only enhances the segmentation performance of the backbone network but also enables it to learn and provide well-calibrated uncertainty estimation.

\subsection{Uncertainty-aware filtering\label{M_4}} 
To apply DEviS to clinical tasks to screen out reliable data, we propose a Uncertainty award filtering strategy (Fig.~\ref{F_1} \textbf{c}). This will benefit the predictive accuracy of ID and OOD data. According to~\cite{2017ece}, a well-calibrated with confidence reflects the reliability of the model. Inspired by this, we first formulate an Uncertainty-based Calibration Error (UCE) to judge the reliability of the data according to the accuracy and uncertainty of the data. For a given validation set ${\mathbb{V}} = \left\{ {{x_1}, \ldots ,{x_m}} \right\}$ and its uncertainty estimation ${{\mathbb{U}}_{\mathbb{V}}} = \left\{ {{\bf{\tilde u}}\left( {{x_1}} \right), \ldots ,{\bf{\tilde u}}\left( {{x_m}} \right)} \right\}$, the UCE of each image ${x_m}$ can be denoted as follows: 
\begin{equation}
\label{E_10}
{\rm{UCE}}\left( {{x_m}} \right) = \left| {acc\left( {{x_m}} \right) - \exp \left( { - {\bf{\tilde u}}\left( {{x_m}} \right)} \right)} \right|,
\end{equation}
where $acc(x_m)$ and $\exp \left( { - U\left( {{x_m}} \right)}\right)$ denote the accuracy and confidence of the $m$-th image data. The ${\bf{\tilde u}}\left(  \cdot  \right)$ represents the mean value of the uncertainty estimate corresponding to the ground truth of the medical image segmentation. To automatically determine the uncertainty threshold, we select the uncertainty value corresponding to the sample with the maximum UCE in the validation set as the threshold. This threshold represents the maximum tolerable uncertainty level associated with the largest acceptable calibration error observed in the validation data. For any test sample, if its uncertainty exceeds this threshold, it is classified as "unreliable"; otherwise, it is considered "reliable". As such, we consider the largest UCE sample to be the tolerable maximum for ID data data, expressed as follows: 
\begin{equation}
\label{E_11}
{\bf{u}}_m^* = {\bf{\tilde u}}\left( {\mathop {\arg \max }\limits_{{x_m} \in {\mathbb{V}}} {\rm{UCE}}\left( {{x_m}} \right)} \right).
\end{equation}
Therefore, we derive the adaptive uncertainty threshold ${\bf{U}}_m^*$ from the uncertainty distribution of the validation dataset ${\mathbb{V}}$, ensuring the selection of the most reliable threshold. For any OOD image $x_{ood}$ of segmentation task, the reliability of the data can be judged according to the following formula:
\begin{equation}
\label{E_12}
{x_{ood}} = \left\{ {\begin{array}{*{20}{c}}
{0,{\bf{\tilde u}}\left( {{x_m}} \right) \ge {\bf{u}}_m^*}\\
{1,{\bf{\tilde u}}\left( {{x_m}} \right) < {\bf{u}}_m^*}
\end{array}} \right.
\end{equation}
$0$ and $1$ denote the unreliable data and reliable data, respectively. It should be noted that for OOD data, the model is likely to filter the data it considers reliable or not. Unreliable data may still require further diagnosis by clinical experts. This thresholding method, based on the statistical distribution of the validation set, avoids the subjectivity and arbitrariness of manual threshold setting, enabling an adaptive and generalizable selection process. It naturally adjusts across different datasets and model architectures, effectively helping the model identify uncertain or OOD samples, thereby supporting further expert review of abnormal data in clinical practice and enhancing the overall safety and robustness of the system.

\subsection{Evidential calibrated training loss function\label{M_5}} 
Due to the imbalance of organ/tumor, our DEviS is first trained with cross-entropy loss function, which is defined as: 
\begin{equation}
\label{E_13}
{{\mathcal L}_{ce}}\left( {\bf{p}}_x,{\bf{y}}_x \right) = \sum\limits_{c = 1}^C { - {\bf{y}}_x^c\log \left( {\bf{p}}_x^c \right)} 
\end{equation}
For simplicity, $ {\bf{y}}_x $ and $ {\bf{p}}_x $ denote the label and predicted probability matrix, respectively, of a random image sample $ x $. The subscript $x $ includes all pixel indices $ i $, $ j $, and $ k $, and the same meaning applies to the subscript $ x $ in subsequent equations. Then, SL associates the Dirichlet distribution with the belief distribution under the framework of evidence theory for obtaining the probability of different classes and uncertainty of different voxels based on the evidence collected from backbone.~Therefore, Eq.~\ref{E_13} can be further refined, resulting in the improved cross-entropy loss (ice) as follows:
\begin{equation}
\label{E_14}
\begin{array}{l}
{{\mathcal L}_{ice}}\left( {{\bf{b}}_x,{\boldsymbol{\alpha}} _x,{\bf{y}}_x} \right)\\
 = \int {\left[ {\sum\limits_{c = 1}^C { - {\bf{y}}_x^c\log ({\bf{b}}_x^c)} } \right]} \frac{1}{{B\left( {{{\boldsymbol{\alpha }}_x}} \right)}}\prod\limits_{c = 1}^C {{{\left( {{\bf{b}}_x^c} \right)}^{{{\alpha }}_x^c - 1}}d{{\boldsymbol{b}}_x}}  \\
 =  - \sum\limits_{c = 1}^C {{\bf{y}}_x^c\left[ {\int {\log ({\bf{b}}_x^c)\frac{1}{{B\left( {{{\bf{\alpha }}_x}} \right)}}\prod\limits_{c = 1}^C {{{\left( {{\bf{b}}_x^c} \right)}^{\alpha _x^c - 1}}d{{\bf{b}}_x}} } } \right]} \\
= \sum\limits_{c = 1}^C {{\bf{y}}_x^c\left( {\psi \left( {{{\bf{S}}_x}} \right) - \psi \left( {{\boldsymbol{\alpha}}_x^c} \right)} \right)} 
\end{array},
\end{equation}
where $\psi \left(  \cdot  \right)$ denote the $digamma$ function. For simplicity, ${{\boldsymbol{b}}_x}$ is the class assignment probabilities on a simplex. Here, we utilize the property of the Dirichlet distribution integral, which is
$\mathbb{E}_{{\rm{D}}({\bf{p}}\mid {\bf{\alpha }})}\left[ {\log ({p^c})} \right] = \psi ({\alpha ^c}) - \psi (S)$, where $S = \sum\nolimits_{c = 1}^C {{\alpha ^c}}$. To guarantee that incorrect labels will yield less evidence, even shrinking to 0, the KL divergence loss function is introduced. We first consider two Dirichlet distributions $D({{\boldsymbol{p}}_x}\mid {{\boldsymbol{\alpha }}_x})$ and $D({{\boldsymbol{p}}_x}\mid {{\boldsymbol{\beta }}_x})$ as below:
\begin{equation}
\label{E_15_pre}
\begin{array}{l}
{{\mathcal L}_{KL}}\left( {{\boldsymbol{\alpha}} _x,{\boldsymbol{\beta}} _x} \right)  \\ 
= \log \left( {\frac{{\Gamma \left( {\sum\nolimits_{c = 1}^C {{\boldsymbol{\alpha}} _x^c} } \right)}}{{\prod\nolimits_{c = 1}^C {\Gamma \left( {{\boldsymbol{\alpha}} _x^c} \right)} }}\frac{{\prod\nolimits_{c = 1}^C {\Gamma \left( {{\boldsymbol{\beta}} _x^c} \right)} }}{{\Gamma \left( {\sum\nolimits_{c = 1}^C {{\boldsymbol{\beta}} _x^c} } \right)}}} \right)\\
 + \sum\nolimits_{c = 1}^C {\left( {{\boldsymbol{\alpha }}_x^c - {\boldsymbol{\beta}} _x^c} \right)} \left[ {\psi \left( {\sum\nolimits_{c = 1}^C {{\boldsymbol{\alpha}} _x^c} } \right) - \psi \left( {\sum\nolimits_{c = 1}^C {{\boldsymbol{\beta}} _x^c} } \right)} \right]
\end{array},
\end{equation}
where $\Gamma \left(  \cdot  \right)$ is the $gamma$ function. For $C$-class classification,  given the label ${\bf{y}}_x$, the predicted Dirichlet distribution ${\boldsymbol{\alpha}}_x $ is expected to be concentrated at the vertex corresponding to the label. This indicates that all Dirichlet parameters should ideally approach~\textbf{1}, except for the parameter linked to the correct label. Accordingly, the aforementioned equation can be rewritten as:
\begin{equation}
\label{E_15}
\begin{array}{l}
{{\mathcal L}_{KL}}\left( {{\boldsymbol{\alpha}} _x},\bf{1} \right)\\
= \log \left( {\frac{{\Gamma \left( {\sum\nolimits_{c = 1}^C {{{\tilde {\boldsymbol{\alpha}} }_x^c}} } \right)}}{{\Gamma (C)\prod\nolimits_{c = 1}^C {\Gamma \left( {{{\tilde {\boldsymbol{\alpha}} }_x^c}} \right)} }}} \right) \\
+ \sum\nolimits_{c = 1}^C {\left( {{{\widetilde {\boldsymbol{\alpha }}}_x^c} - 1} \right)} \left[ {\psi \left( {{{\widetilde {\boldsymbol{\alpha }}}_x^c}} \right) - \psi \left( {\sum\nolimits_{c = 1}^C {{{\widetilde {\boldsymbol{\alpha }}}_x^c}} } \right)} \right]
\end{array},
\end{equation}
Where ${{\tilde {{{\boldsymbol{\alpha}} }} }_x^c} = {{\bf{y}}_x^c} + \left( {{\bf{1}} - {{\bf{y}}_x^c}} \right) \odot {{{{\boldsymbol{\alpha}} }} _x^c}$ denotes the adjusted parameters of the Dirichlet distribution, which is used to ensure that ground-truth class evidence is not mistaken for 0. We simplify this equation as $ {{\mathcal L}_{KL}}\left( {{\boldsymbol{\alpha}} _x} \right) $. More detailed derivations for Eq.~\ref{E_14} and Eq.~\ref{E_15} can be found in~\cite{2020trusted}. Therefore, the evidential deep learning process can be defined as:
\begin{equation}
\label{E_16}
{{\cal L}_{E}}\left( {{\bf{b}}_x,{\boldsymbol{\alpha}} _x,{\bf{y}}_x} \right) = {{\cal L}_{ice}}\left( {{\bf{b}}_x,{\boldsymbol{\alpha}} _x,{\bf{y}}_x} \right) + {\lambda}{{\cal L}_{KL}}\left( {\boldsymbol{\alpha}} _x \right).
\end{equation}
where ${\lambda}$ is the balance annealing coefficient, defined as $\lambda = \min(1, \lambda_{\text{ini}}) \in [0,1]$, where the initial value is set as $\lambda_{\text{initial}} = t/50$, which equals 0.02 when $t=1$. Furthermore, the Dice score is an important metric for judging the performance of organ/tumor segmentation. Therefore, we use a Dice loss ${{\cal L}_{Dice}}\left( {{{\bf{b}}_x},{{\bf{y}}_x}} \right)$. Then, according to Sec.~\ref{M_3} of calibrated uncertainty penalty, the loss function for well-calibrated uncertainty can be defined as follows:
\begin{equation}
\label{E_19}
\begin{array}{*{20}{c}}
\begin{aligned}
&{{\mathcal L}_{CUP}}\left( {{{\bf{b}}_x},{{\bf{u}}_x},{{\bf{y}}_x}} \right) \\
&=  - {\beta _t}\sum\limits_{c = 1}^C {{\bf{b}}_x^c\log \left( {1 - {{\bf{u}}_x}} \right)}  - \left( {1 - {\beta _t}} \right)\sum\limits_{c = 1}^C {\left( {1 - {\bf{b}}_x^c} \right)\log \left( {{{\bf{u}}_x}} \right)}.
\end{aligned}
\end{array}
\end{equation}
To guide the model optimization at the early stage of network training, $\left({1 - {\beta _t}}\right)$ is noted as the annealing factor, which is defined by ${\beta _t}{\rm{ = }}{\beta _0}{e^{\left\{ { - ({\rm{In}}{\beta _0})t/T} \right\}}}$. $T$ and $t$ are the total epochs and the current epoch, respectively. The value of ${\beta _0}$ is set to 0.01.~Finally, the overall evidential calibrated training loss function of our proposed network can be defined as follows:
\begin{equation}
\label{E_20}
\begin{array}{*{20}{c}}
\begin{aligned}
&{\mathcal L}\left( {{{\bf{b}}_x},{\boldsymbol{\alpha}} _x, {{\bf{u}}_x}, {{\bf{y}}_x}} \right)\\
&= {{\mathcal L}_{E}}\left( {{{\bf{b}}_x},{\boldsymbol{\alpha}} _x,{{\bf{y}}_x}} \right) +  {{\mathcal L}_{Dice}}\left( {{{\bf{b}}_x},{{\bf{y}}_x}} \right) + {{\mathcal L}_{CUP}}\left( {{{\bf{b}}_x},{{\bf{u}}_x},{{\bf{y}}_x}} \right)
\end{aligned}
\end{array}
\end{equation}
 Details of the training process of the proposed DEviS method, together with further ablation studies on ${\lambda _{{\rm{ini}}}}$, $W$, the calibrated method $e^{-u}$ in comparison with the Softmax function, the generalization ability of CUP, and the effect of different uncertainty thresholds, are presented in the Supplementary Materials.

\section{Experimental results}
\subsection{Experimental Setup \& Datasets} 
1) Experimental Setup: Our proposed network is implemented in PyTorch and trained on NVIDIA GeForce RTX 2080Ti. We adopt the Adam to optimize the overall parameters. The initial learning rates for different datasets are set to be 0.0002 (ISIC2018) and 0.002 (BraTS2019). The poly learning strategy is used by decaying each iteration with a power of 0.9. The maximum of the epoch is set to 200. The batch sizes for the lesion segmentation, and brain tumor segmentation are set to 16 and 2. All the following experiments adopted a five-fold cross-validation strategy to prevent performance improvement caused by accidental factors. For the ISIC2018 dataset, we used the data augmentation by random cropping, flipping, and random rotation as same as~\cite{2020CA}. For the BraTS2019 dataset, the data augmentation techniques are similar as~\cite{TransBTS}. For the clinical application of OOD detection, we conducted the experiments on the Johns Hopkins OCT dataset and the Duke OCT dataset with DME. In the application, the initial learning rate for the dataset are set to be 0.0001. The poly learning strategy is used by decaying each iteration with a power of 0.9. The maximum of the epoch is set to 100. The batch sizes is set to 8. The following will provide additional detailed information about the ISIC2018 and BraTS2019 datasets used, as well as information regarding degraded image data. It also elaborates on the datasets utilized in clinical applications.

2) Datasets: In this study, we evaluate our method across a broad range of publicly available 2D and 3D medical image segmentation benchmarks. These include the ISIC2018~\cite{ISIC2018} dataset for skin lesion segmentation, KiTS2021~\cite{kits21} for kidney and tumor segmentation in abdominal CT, LiTS2017~\cite{HDenseU} for liver tumor segmentation, and BraTS2019~\cite{BraTSbench} for multi-modality brain tumor segmentation. To ensure robustness under OOD conditions, test sets were augmented with noise, blur, and random masking. Beyond these benchmarks, we further validated the method on ophthalmic datasets, including the Johns Hopkins OCT(\url{https://iacl.ece.jhu.edu/index.php?title=Main_Page}) dataset (healthy and multiple sclerosis cases) and the Duke-OCT-DME(\url{https://people.duke.edu/~sf59/Chiu_BOE_2014_dataset.htm}) dataset (diabetic macular edema), as well as retinal quality and vessel segmentation datasets (FIVES~\cite{Fives2022} and DRIVE~\cite{DRIVE2013}). Finally, cardiac segmentation performance was examined on the ACDC2018~\cite{bernard2018deep} dataset. More detailed dataset descriptions, preprocessing procedures, and OOD settings are provided in the Supplementary Materials.

\subsection{Compared Methods \& Evaluation\label{M_6}} 
1) Compared Methods: In this study, we seamlessly integrated our proposed method into both U-Net-based and transformer-based frameworks to rigorously evaluate its robustness and reliability pre and post-integration. Comparative analyses were conducted against state-of-the-art methods employing uncertainty estimation techniques. The U-Net variants encompassed traditional U-Net~\cite{Unet15}, V-Net~\cite{VNet}, Attention-UNet~\cite{AttentionU18}, and nnU-Net~\cite{isensee2021nnu}, while the transformer-based approach was predominantly represented by nnFormer~\cite{zhou2023nnformer}. We also employed the Vision Mamba-based method, VM-UNet~\cite{zhang2024vm}, for the 2D medical image segmentation task. Additionally, our approach underwent comprehensive comparisons with uncertainty estimation methodologies, including U-Net-based Monte Carlo dropout sampling (DU)~\cite{17dropoutCV}, U-Net-based ensemble (UE)~\cite{ensemble17}, Probabilistic-UNet (PU)~\cite{2018probabilisticU}, Bayesian QuickNAT (BQNAT)~\cite{roy2019bqnat}, Test-Time Augmentation (TTA)~\cite{wang2018TTA} and our conference version~\cite{zou2022tbrats}. These comparisons were undertaken to elucidate the superiorities and limitations of our proposed method in the context of existing cutting-edge techniques. For the semi-supervised experiments, the Mean Teacher (MT)~\cite{tarvainen2017mean} method is considered the most classical baseline. To demonstrate the generality of DEviS, we integrate it into the MT framework, denoted as DEviS-MT. Additionally, we compare our approach with two representative uncertainty-based semi-supervised segmentation methods: Expectation Maximization (EM)~\cite{vu2019advent}, Uncertainty-Aware Mean Teacher (UA-MT)~\cite{yu2019uncertainty} and uncertainty rectified pyramid consistency(URPC)~\cite{luo2022semi}.

2) Evaluation Metrics: The following metrics are employed for quantitative evaluation. (a) The Dice score (Dice) and (b) Average symmetric
surface distance (ASSD) is adopted as an intuitive evaluation of segmentation accuracy. Given the absence of ground truth for the uncertainty estimate, we utilize the same evaluation metrics as referenced in ~\cite{guo2017calibration,2019assessing} to assess its performance. (c) Expected calibration error (ECE) \cite{guo2017calibration,2019assessing} and (d) Uncertainty-error overlap (UEO) \cite{guo2017calibration,2019assessing} are used as evaluation of uncertainty estimations. For the semi-supervised methods, following~\cite{luo2022semi}, we adopt Dice coefficient, ASSD, and HD95 as evaluation metrics.

\subsection{DEviS improves robustness and calibration for different base networks.}
To illustrate this, we tackled three challenging medical image segmentation tasks using different datasets: (1) 2D skin lesion images from the ISIC2018 dataset, and (2) multi-modal 3D MRI images from the BraTS2019 dataset. These tasks encompass the segmentation of various tissues and tumors, aiming to achieve accurate and robust segmentation performance across different levels of normal and degraded (noise, blur, and random mask) conditions in medical images, alongside reliable and efficient uncertainty estimation.

\subsubsection{DEviS for 2D medical image}
We initially enhance the robustness and accuracy of the base network in skin lesion boundary segmentation~\cite{ISIC2018, 2020CA} using DEviS. We conducted studies at different levels of Gaussian noise and random masking based on the ISIC2018 dataset (2D dermoscopic image) to validate the robustness of our proposed framework.

\begin{figure*}
\centering
\includegraphics[width=1\linewidth]{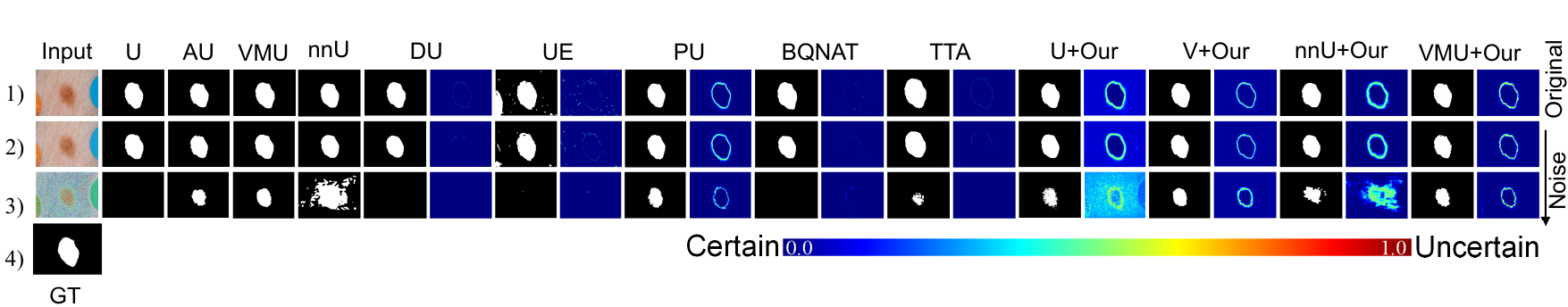}
\caption{visual comparison results of different methods on the ISIC 2018 dataset with different methods: 1) original input and its results; 2-3) input with Gaussian noise ($\sigma=0.1, 0.5$) and its results; 4) Ground truth.}
\label{F_isic}
\end{figure*}

As shown in the Tab.~\ref{Tab_isic1}, the comparison results between DEviS and other U-Net variants under different Gaussian noise levels ratios are presented. Tab.~\ref{Tab_isic1} indicates a gradual degradation in performance for U-Net, AU-Net, V-Net, nnU-Net, and VM-UNet, particularly at higher mask ratios and noise levels. Upon applying DEviS, the results exhibit a certain degree of robustness to interference. When equipped with DEviS, U-Net demonstrates an average improvement of 10.6\% and 8.9\% in Dice metric under degraded conditions of Gaussian noise and random masking, respectively. Additionally, the generated uncertainty estimates, as illustrated in Fig.~\ref{F_isic}, can be utilized by researchers and clinicians to discern the unreliability of the data.

As shown in Tab.~\ref{Tab_isic2}, the comparison results of the ECE and UEO metrics between the proposed method and other uncertainty estimation methods are presented. It reveales that BQNAT, DU, PU, UE, and TTA methods were significantly affected by noise and masking, while the perturbation on U-Net, V-Net, nnU-Net, and VM-UNet methods was relatively minor after applying DEviS. A comparison of uncertainty estimation results using ECE and UEO metrics indicated that U-Net, V-Net, nnU-Net, and VM-UNet with DEviS achieved better uncertainty estimation. Visualizations of segmentation results and uncertainty estimation as shown in Fig.~\ref{F_isic}, demonstrate that the proposed DEviS method provides more reliable uncertainty estimation for target edges and the noised pixels.

\begin{table}[htbp]
  \centering
  \small
  \caption{The quantitative comparisons with different backbone based methods on the ISIC2018 dataset under differing Gaussian noise: Dice and ASSD metrics with different Gaussian noise $\sigma=\left\{ {0,0.1,0.3,0.5} \right\}$}
    \resizebox{0.5\textwidth}{!}{
    \begin{tabular}{lcccccccc}
    \toprule
    \multicolumn{1}{c}{\multirow{2}[4]{*}{Method}} & \multicolumn{4}{c}{Dice ($\sigma =$)}      & \multicolumn{4}{c}{ASSD ($\sigma =$)} \\
\cmidrule{2-9}          & 0     & 0.1   & 0.3   & 0.5   & 0     & 0.1   & 0.3   & 0.5 \\
    \midrule
    U~\cite{Unet15}    & 82.82  & 79.20  & 50.07  & 6.66  & 8.78  & 10.94  & 26.77  & 51.64  \\
    V~\cite{VNet}    & 86.31  & 84.76  & 75.12  & 54.74  & 7.15  & 8.46  & 13.98  & 23.93  \\
    AU~\cite{AttentionU18}    & 85.49  & 81.77  & 50.49  & 9.62  & 7.56  & 9.51  & 25.37  & 45.01  \\
    nnU~\cite{isensee2021nnu}   & 86.28  & 85.04  & 70.37  & 48.94  & 6.72  & 7.18  & 13.63  & 28.10  \\
    VM-UNet~\cite{zhang2024vm} & 85.82  & 84.64  & 70.02  & 51.31  & 7.75  & 8.16  & 17.82  & 22.67  \\
    U+Our & 86.80  & 82.18  & 60.72  & 28.05  & 7.41  & 8.99  & 21.06  & 37.62  \\
    V+Our & 86.57  & 85.38  & 78.11  & 59.83  & 6.71  & 7.50  & 15.22  & 19.88  \\
    nnU+Our & 87.31  & \textbf{86.75}  & \textbf{78.88}  & \textbf{62.31}  & \textbf{6.25}  & \textbf{6.64}  & \textbf{10.53}  & \textbf{18.38}  \\
    VM-UNet+Our & \textbf{87.78}  & 85.52  & 72.64  & 53.35  & 7.50  & 8.06  & 15.77  & 20.89  \\
    \bottomrule
    \end{tabular}%
    }
  \label{Tab_isic1}%
\end{table}%

\begin{table}[htbp]
  \centering
  \caption{The quantitative comparisons with uncertainty estimation based methods on the ISIC2018 dataset under differing Gaussian noise: ECE and UEO metrics with different Gaussian noise $\sigma=\left\{ {0,0.1,0.3,0.5} \right\}$. he numbers in the table need to be multiplied by ($0.01 \times $).}
  \resizebox{0.5\textwidth}{!}{
    \begin{tabular}{lcccccccc}
    \toprule
    \multicolumn{1}{c}{\multirow{2}[4]{*}{Method}} & \multicolumn{4}{c}{ECE ($0.01 \times $, $\sigma =$)}      & \multicolumn{4}{c}{UEO ($0.01 \times $, $\sigma =$)} \\
\cmidrule{2-9}          & 0     & 0.1   & 0.3   & 0.5   & 0     & 0.1   & 0.3   & 0.5 \\
    \midrule
    PU~\cite{2018probabilisticU}    & 5.54  & 7.11  & 12.99 & 19.40  & 86.81 &  83.56 & 66.45 & 29.50 \\
    UE~\cite{ensemble17}    & 4.91  & 6.33  & 14.17 & 19.22 & 85.87 & 83.74 & 57.24 & 37.44 \\
    DU~\cite{17dropoutCV}    & 5.24  & 6.87  & 16.32 & 20.86 & 86.57 &  83.17 & 54.37 & 35.59 \\
    BQNAT~\cite{roy2019bqnat} & 5.21  & 6.86  & 15.95 & 21.50  & 86.35 &  83.36 & 54.31 & 32.86 \\
    TTA~\cite{wang2018TTA}   & 4.77  & 6.50   & 15.91 & 18.17 & 88.30  & 79.68 & 50.61 & 39.91 \\
    U+Our & 4.73  & 6.22  & 9.14  & 12.14 & 87.44 & 83.81 & 77.49 & 67.4 \\
    V+Our & 5.29  & 6.20   & 9.67  & 14.80  &90.27 &  89.33 & 83.56 & 70.30 \\
    nnU+Our & 4.69  & 5.24  & 8.98  & 15.28 &90.98 &  \textbf{90.88} & \textbf{85.26} & \textbf{72.43} \\
    VM-UNet+Our & \textbf{0.98} &  \textbf{1.53}  & \textbf{4.91}  & \textbf{9.78}  & \textbf{91.16} & 89.04 & 80.93 & 60.79 \\
    \bottomrule
    \end{tabular}%
    }
  \label{Tab_isic2}%
\end{table}%

\subsubsection{DEviS for 3D medical image}
For 3D medical image segmentation, we validate our model on the KiTS and LiTS datasets. First,  we evaluated the robustness and accuracy of DEviS in enhancing both U-Net-based and Transformer-based backbone networks for the diagnosis of kidney diseases. To validate the reliability and robustness of this approach, we conducted experiments on the KiTS2021 dataset (3D CT) under Gaussian noise (with a standard deviation of 0.1) and Gaussian blur (with a variance of 7 and a kernel size of $(7,7)$), aiming to achieve robust medical image segmentation.

\begin{figure}
\centering
\includegraphics[width=1\linewidth]{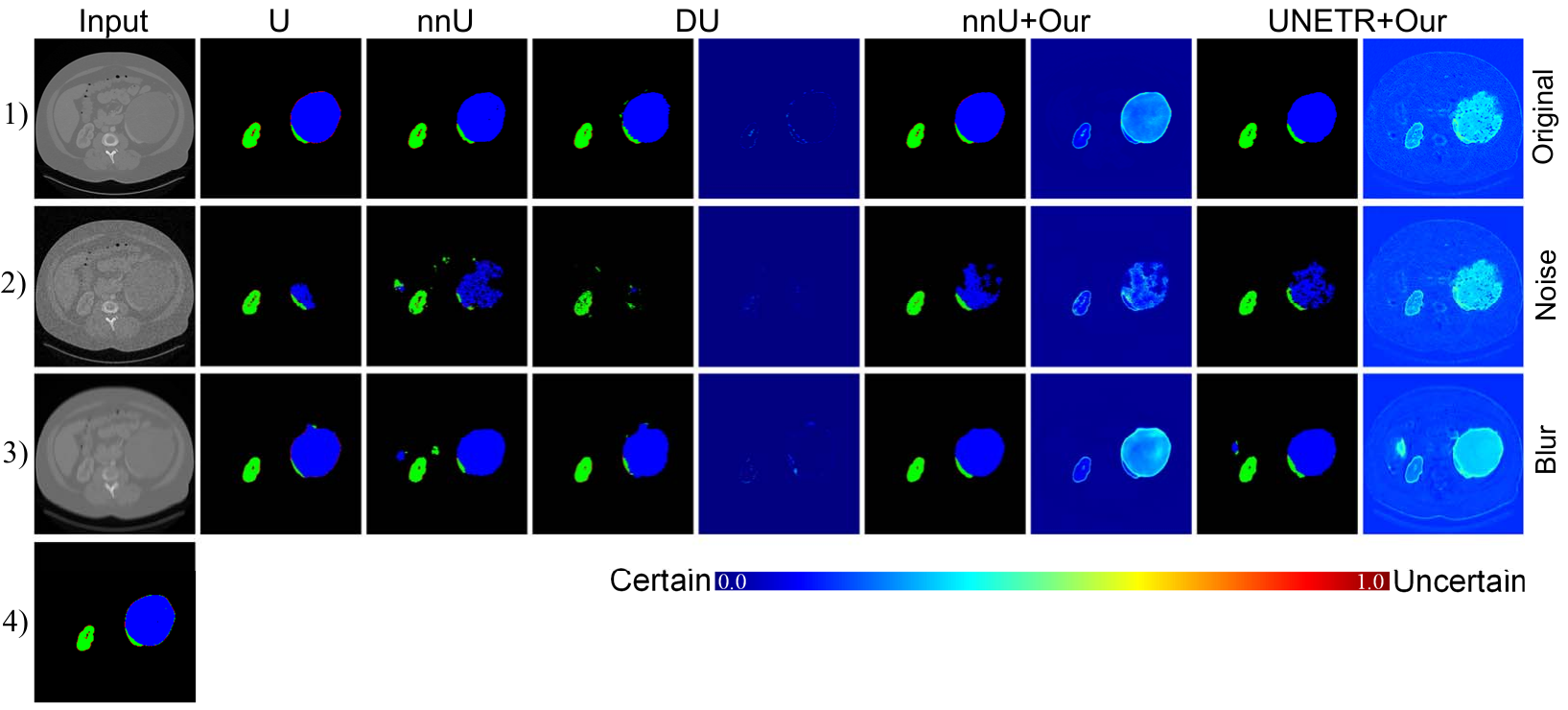}
\caption{visual comparison results of different methods on the KiTS2021 dataset with different methods: 1) original input and its results; 2-3) input with Gaussian noise (0.1) and Gaussian blur ($\sigma$=7, k = 7) and its results; 4) Ground truth.}
\label{F_kits}
\end{figure}

As shown in table~\ref{tab_kits1}, as the degradation conditions worsen, kidney segmentation performance in terms of Dice and ASSD metrics shows a more significant decline for both U-Net-based and Transformer-based methods. However, this trend can be mitigated by applying the DEviS approach. Taking UNETR as an example, after integrating the DEviS model under two degradation conditions, robustness is significantly improved, with an average Dice score increase of 9.98\% and 4.47\%, respectively. Further, as shown in the table~\ref{tab_kits2}, we compare our proposed method with the Dropout-based uncertainty estimation method, DU. The results indicate that our method outperforms DU, particularly in terms of the ECE and UEO metrics. Additionally, we present the visual comparison results of Kidney segmentation, as shown in the figure~\ref{F_kits}. It can be observed that the baseline models, nnU-Net and UNETR, equipped with our method, achieve more robust visual results. Similarly, as illustrated in Figure~\ref{F_kits}, the uncertainty estimation maps generated by the DEviS framework can assist clinicians in diagnosis and analysis. Furthermore, we evaluated the reliability and calibration capability of the method using ECE and UEO under the same conditions. The results demonstrate that the DEviS approach outperforms the DU method significantly.

\begin{table}[htbp]
  \centering
  \caption{The quantitative comparisons with uncertainty estimation based methods on the KiTS2021 dataset under differing Gaussian noise and blur: Dice and ASSD metrics.}
    \begin{tabular}{lrrrrrr}
    \toprule
    \multicolumn{1}{c}{\multirow{2}[4]{*}{Method}} & \multicolumn{2}{c}{Normal} & \multicolumn{2}{c}{Noise (0.1)} & \multicolumn{2}{c}{Blur (7,7)} \\
\cmidrule{2-7}          & \multicolumn{1}{c}{Dice} & \multicolumn{1}{c}{ASSD} & \multicolumn{1}{c}{Dice} & \multicolumn{1}{c}{ASSD} & \multicolumn{1}{c}{Dice} & \multicolumn{1}{c}{ASSD} \\
    \midrule
    nnU~\cite{isensee2021nnu}   & 93.77  & \textbf{1.07}  & 76.28  & 6.05  & 77.19  & 5.73  \\
    UNETR~\cite{2022unetr} & 85.95  & 4.08  & 68.16  & 6.97  & 73.62  & 6.40  \\
    DU~\cite{17dropoutCV}   & 83.07  & 4.32  & 36.67  & 9.94  & 67.24  & 7.34  \\
    nnU+Our & \textbf{95.02}  & 1.22  & \textbf{82.89}  & \textbf{4.20}  & \textbf{82.92}  & \textbf{4.24}  \\
    UNETR+Our & 86.77  & 4.52  & 78.14  & 5.76  & 78.09  & 5.41  \\
    \bottomrule
    \end{tabular}%
  \label{tab_kits1}%
\end{table}%

\begin{table}[htbp]
  \centering
  \caption{The quantitative comparisons with uncertainty estimation based methods on the KiTS2021 dataset under differing Gaussian noise and blur: ECE and UEO metrics. The numbers in the table need to be multiplied by ($0.01 \times $).}
    \begin{tabular}{lcccccc}
    \toprule
    \multicolumn{1}{c}{\multirow{2}[4]{*}{Method}} & \multicolumn{2}{c}{Normal} & \multicolumn{2}{c}{Noise (0.1)} & \multicolumn{2}{c}{Blur (7, 7)} \\
\cmidrule{2-7}          & ECE   & UEO   & ECE   & UEO   & ECE   & UEO \\
    \midrule
     DU~\cite{17dropoutCV}   & 0.60  & 83.37  & 4.41  & 19.60  & 2.17  & 67.55  \\
    nnU+Our & \textbf{0.12}  & \textbf{95.82}  & 1.48  & \textbf{86.20}  & \textbf{0.56}  & \textbf{90.61}  \\
    UNETR+Our & 0.32  & 87.47  & \textbf{1.32}  & 79.30  & 1.07  & 79.08  \\
    \bottomrule
    \end{tabular}%
  \label{tab_kits2}%
\end{table}%

b) We further tasked DEviS with enhancing the robustness and accuracy of the base networks in diagnosing hepatocellular diseases~\cite{HDenseU}. To verify the reliability and robustness of our method, we conducted studies with the LiTS2017 dataset (3D CT) under differing levels of Gaussian noise (with a standard deviation of 0.4), Gaussian blur (with a variance of 15 and a kernel size of 20),  and random masking (utilizing an 8-pixel mask size and a mask ratio of 0.1) to achieve robust medical image segmentation.

As shown in the Tab.~\ref{tab_lits1}, as the degradation conditions worsened, the liver segmentation results based on U-Net methods exhibited a more pronounced decline in Dice and ASSD metrics. However, this trend could be mitigated by applying the DEviS method. Taking nnU-Net as an example, the robustness was significantly enhanced after integrating the DEviS model under three degradation conditions, with average improvements of 23.7\%, 8.6\%, and 7.0\% in Dice metrics. Similarly, as depicted in Fig.~\ref{F_livsbrats} \textbf{a}, the uncertainty estimation maps generated by the integrated DEviS method can assist physicians in diagnosis and analysis.

We repeated the study with uncertainty-based algorithms to further compare the calibrated uncertainty for segmentation. Performance in uncertainty estimation for most uncertainty-based methods degrades with increasing Gaussian noise, Gaussian blur, and random masking (Tab.~\ref{tab_lits2}). However, the backbones equipped with DEviS mitigates this degradation. Under normal conditions, our framework provides higher uncertainty for liver edges compared to other methods. Under noised conditions, our framework is more sensitive to unseen regions and provides high uncertainty (Fig.~\ref{F_livsbrats} \textbf{a}). DEviS thus allows clinicians or annotators to better focus on areas of uncertainty.

\begin{table}[htbp]
  \centering
  \caption{The quantitative comparisons with uncertainty estimation based methods on the LiTS017 dataset under differing Gaussian noise, blur and mask: Dice and ASSD metrics.}
  \resizebox{0.5\textwidth}{!}{
    \begin{tabular}{lrrrrrrrr}
    \toprule
    \multicolumn{1}{c}{\multirow{2}[4]{*}{Method}} & \multicolumn{2}{c}{Normal} & \multicolumn{2}{c}{Noise (0.4)} & \multicolumn{2}{c}{Blur (15,20)} & \multicolumn{2}{c}{Mask (0.1)} \\
\cmidrule{2-9}          & \multicolumn{1}{c}{Dice} & \multicolumn{1}{c}{ASSD} & \multicolumn{1}{c}{Dice} & \multicolumn{1}{c}{ASSD} & \multicolumn{1}{c}{Dice} & \multicolumn{1}{c}{ASSD} & \multicolumn{1}{c}{Dice} & \multicolumn{1}{c}{ASSD} \\
    \midrule
    U~\cite{Unet15}     & 85.88  & 4.10  & 44.23  & 7.67  & 80.76  & 5.01  & 71.58  & 5.46  \\
    V~\cite{VNet}     & 92.78  & 1.17  & 34.89  & 12.92  & 85.87  & 2.66  & 54.09  & 6.07  \\
    AU~\cite{AttentionU18}    & 86.91  & 3.73  & 28.57  & 11.09  & 81.57  & 4.69  & 72.91  & 6.90  \\
    nnU~\cite{isensee2021nnu}   & 90.55  & 1.85  & 11.17  & 21.64  & 77.95  & 5.66  & 76.59  & 3.82  \\
    U+Our & 93.30  & 1.27  & \textbf{55.81}  & \textbf{7.09}  & 87.58  & 2.19  & 76.19  & 3.34  \\
    V+Our & \textbf{94.38}  & \textbf{0.93}  & 48.83  & 8.96  & 87.65  & 2.16  & 78.54  & 2.50  \\
    AU+Our & 93.02  & 1.57  & 33.82  & 7.52  & \textbf{87.66}  & 2.84  & 72.08  & 4.10  \\
    nnU+Our & 94.64  & 0.02  & 47.28  & 0.11  & 87.25  & \textbf{2.03}  & \textbf{85.74}  & \textbf{1.57}  \\
    \bottomrule
    \end{tabular}%
    }
  \label{tab_lits1}%
\end{table}%

\begin{table}[htbp]
  \centering
  \caption{The quantitative comparisons with uncertainty estimation based methods on the LiTS2017 dataset under differing Gaussian noise, blur and mask: ECE and UEO metrics. The numbers in the table need to be multiplied by ($0.01 \times $).}
    \resizebox{0.5\textwidth}{!}{
    \begin{tabular}{lrrrrrrrr}
    \toprule
    \multicolumn{1}{c}{\multirow{2}[4]{*}{Method}} & \multicolumn{2}{c}{Normal} & \multicolumn{2}{c}{Noise (0.4)} & \multicolumn{2}{c}{Blur (15,20)} & \multicolumn{2}{c}{Mask (0.1)} \\
\cmidrule{2-9}          & \multicolumn{1}{c}{ECE} & \multicolumn{1}{c}{UEO} & \multicolumn{1}{c}{ECE} & \multicolumn{1}{c}{UEO} & \multicolumn{1}{c}{ECE} & \multicolumn{1}{c}{UEO} & \multicolumn{1}{c}{ECE} & \multicolumn{1}{c}{UEO} \\
    \midrule
    PU~\cite{2018probabilisticU}    & \textbf{1.50}  & 95.12  & 16.78  & 53.03  & 4.31  & 87.48  & 13.00  & 50.38  \\
    UE~\cite{ensemble17}    & 3.06  & 85.77  & 15.01  & 33.47  & 4.97  & 78.46  & 6.88  & 63.62  \\
    DU~\cite{17dropoutCV}    & 3.36  & 87.97  & 17.42  & 18.41  & 4.76  & 82.87  & 10.82  & 61.88  \\
    BQNAT~\cite{roy2019bqnat} & 5.38  & 81.48  & 17.99  & 13.09  & 6.65  & 77.23  & 11.00  & 62.96  \\
    TTA~\cite{wang2018TTA}   & 3.91  & 83.86  & 17.43  & 15.75  & 5.24  & 79.27  & 11.26  & 58.63  \\
    U+Our & 1.97  & 93.53  & \textbf{10.20}  & 67.79  & 3.10  & 89.92  & 6.27  & 83.62  \\
    V+Our & 1.69  & 94.93  & 12.84  & 56.91  & 3.20  & 90.04  & 5.92  & 83.56  \\
    AU+Our & 1.98  & 93.14  & 14.44  & 59.25  & \textbf{3.05}  & 89.03  & 7.42  & 80.15  \\
    nnU+Our & 1.64  & \textbf{95.19}  & 10.64  & \textbf{70.81}  & 3.18  & \textbf{90.06}  & \textbf{4.44}  & \textbf{88.56}  \\
    \bottomrule
    \end{tabular}%
    }
  \label{tab_lits2}%
\end{table}%

\subsubsection{DEviS for Multi-modality 3D medical image}
To further validate the robustness and accuracy of the DEviS model in diagnosing brain tumor diseases, experiments were conducted under differing levels of Gaussian noise, Gaussian blur, and random masking. It is worth mentioning that we included the nnFormer model based on the Transformer architecture, which has shown good performance in 3D medical image segmentation, for comparison in this study.

\begin{figure*}
\centering
\includegraphics[width=1\linewidth]{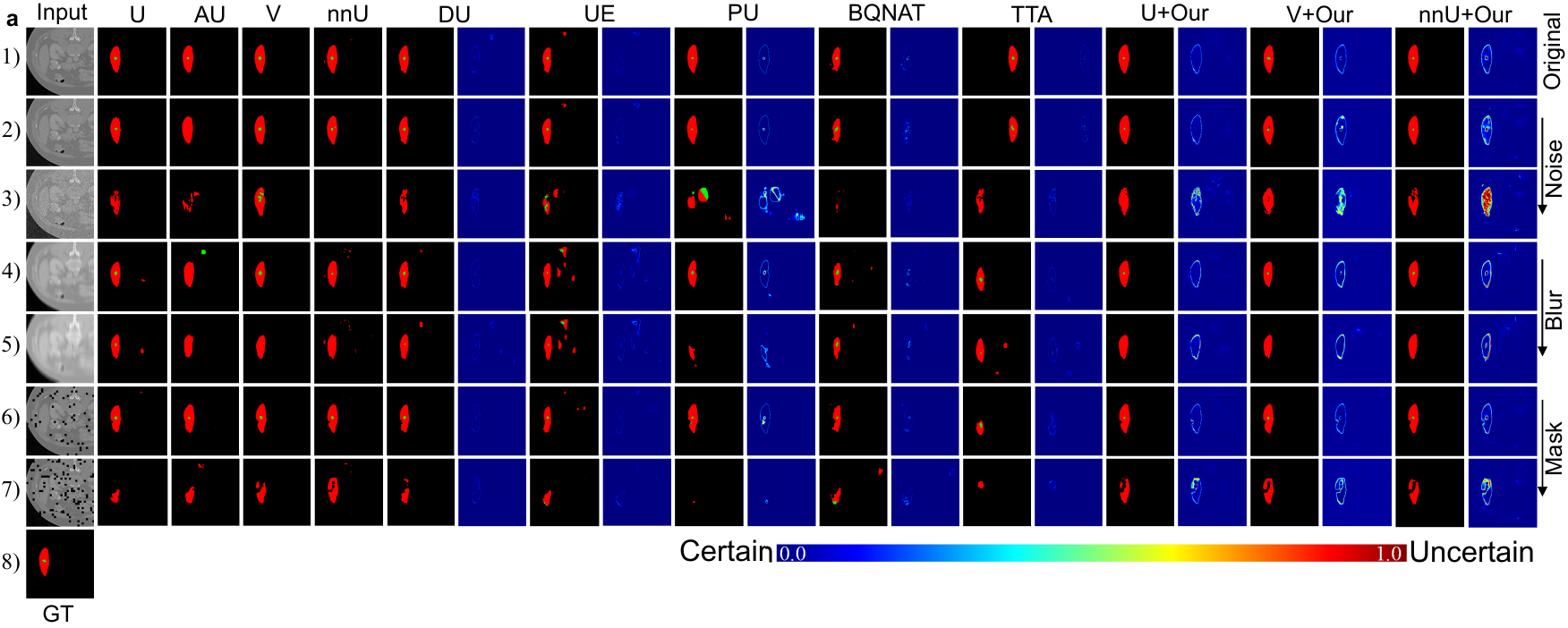}
\includegraphics[width=1\linewidth]{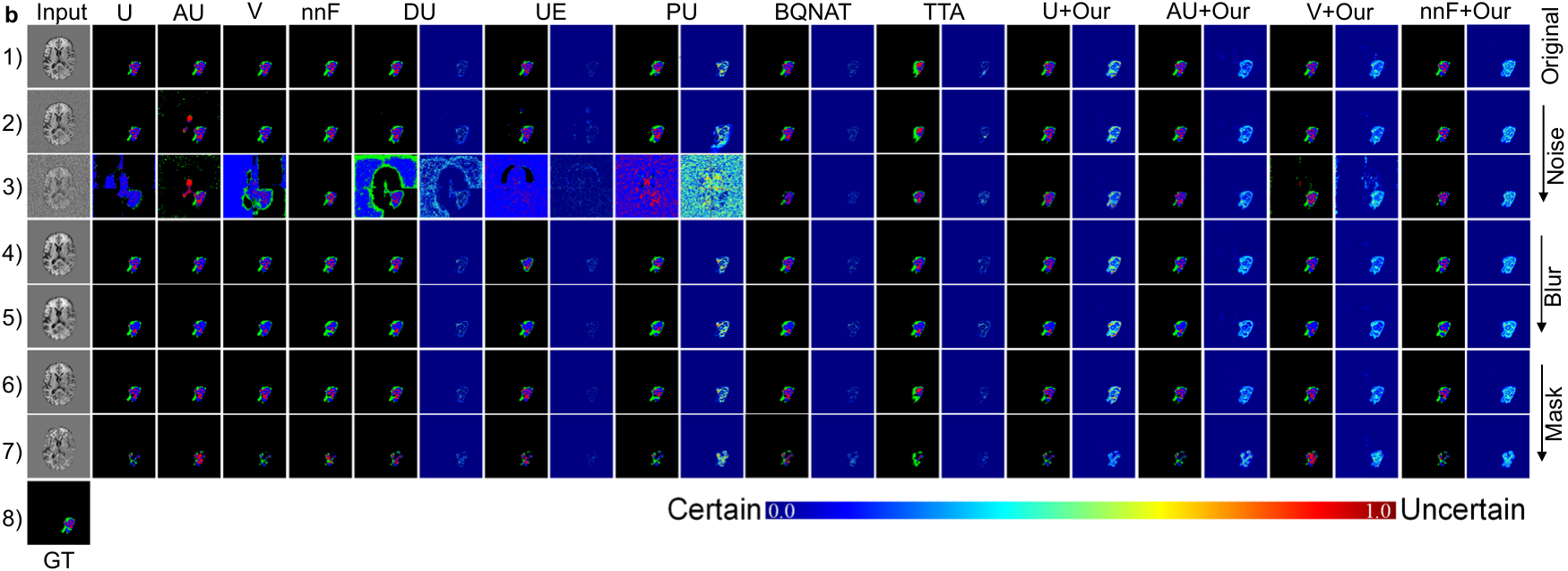}
\caption{\textbf{a.} The Visual comparison of liver segmentation results using different methods on the LiTS2017 dataset: 1) original input and its results; (2-3) Input under Gaussian noise ($\sigma=0.2, 0.4$) and its results; (4-5) Input under Gaussian blur ($\left\{ {\left( {{\sigma},k} \right)} \right\} = \left\{ {\left( {{\rm{15, 10}}} \right),\left( {{\rm{35, 20}}} \right)} \right\}$) and its results;  (6-7) Input under random mask ratio ($\sigma=0.1, 0.3$) and its results; (8) Ground truth. \textbf{b.} The visual comparison of brain tumor segmentation results with different methods. 1) Original input (T1 as an example); 2)-3) Gaussian noise input under ($\sigma=1, 2$) and its results; 4)-5) Input under Gaussian blur ($\left\{ {\left( {{\sigma},k} \right)} \right\} = \left\{ {\left( {{\rm{3, 3}}} \right),\left( {{\rm{7, 7}}} \right)} \right\}$) and its results. 6)-7) Input under random mask ratio $MR=\left\{ {0.1, 0.4} \right\}$ and its results; 8) Ground Truth. }
\label{F_livsbrats}
\end{figure*}

As shown in Tab.~\ref{tab_brats1}, under normal conditions, V-Net, Attention-UNet, nnU-Net, and nnFormer demonstrate comparable performance; however, their segmentation performance begins to degrade under different levels of Gaussian noise, Gaussian blur, and random masking, with the impact becoming more pronounced as the severity increases. The integration of DEviS enhances the robustness under different levels of Gaussian noise, Gaussian blur, and random masking. Taking the Attention-UNet method based on U-Net as an example, the introduction of DEviS results in an average increase of 12.6\% and 5.1\% in the Dice metric under Gaussian noise and Gaussian blur conditions, respectively. Similarly, considering the nnFormer based on the Transformer, the introduction of DEviS leads to an average increase of 2.21\% and 5.93\% in the Dice metric under Gaussian noise and random masking conditions, respectively. It is noteworthy that DEviS exhibits superior robustness in uncertainty estimation compared to the conference version of the TBraTS~\cite{zou2022tbrats} method. Additionally, as depicted in Fig.~\ref{F_livsbrats} \textbf{b}, the segmentation results of different methods show that under Gaussian noise, introducing the DEviS model can slightly improve the segmentation results of nnU-Net and V-Net, while U-Net and Attention-UNet with DEviS demonstrate more robust segmentation results. This is because nnU-Net and V-Net themselves possess certain noise resistance capabilities. We also observed that Attention-UNet exhibited inconsistent performance across the ISIC2018 and BraTS2019 datasets, with our method achieving superior performance when integrated. Our analysis is as follows. First, the ISIC2018 dataset has ambiguous lesion boundaries and high texture variability, making it challenging for the attention mechanism to effectively focus on key regions, resulting in lower Dice scores. Second, in the BraTS2019 dataset, the tumor structures are relatively well-defined, allowing the attention mechanism to better capture different lesion regions, leading to a more significant improvement with AU+our. Finally, our method integrates both evidential uncertainty and calibrated uncertainty penalization, which further optimizes attention distribution, particularly in complex multi-class segmentation tasks. Further analyses of challenging failure cases of uncertainty estimation on ISIC2018, LiTS2017, and BraTS2019 are provided in the Supplementary Materials.

\begin{table*}[htbp]
  \centering
  \caption{The quantitative comparisons with different backbone based methods on the BraTS2019 dataset under differing Gaussian noise: Dice and ASSD metrics with different Gaussian noise $\sigma=\left\{ {0,0.5,1.0,1.5} \right\}$, Gaussian blur ($\left\{ {\left( {{\sigma},k} \right)} \right\} = \left\{ {\left( {{\rm{3, 3}}} \right),  {\left( {{\rm{5, 5}}} \right)} ,\left( {{\rm{7, 7}}} \right)} \right\}$ and different mask ratio ($MR=\left\{ {0.1, 0.25, 0.4} \right\}$)). }    
  \resizebox{0.95\textwidth}{!}{
  \begin{tabular}{lrrrrrrrrrrrrrrrrrrrr}
    \toprule
    \multicolumn{1}{c}{\multirow{3}[6]{*}{Method}} & \multicolumn{2}{c}{Normal} & \multicolumn{6}{c}{Noise}                     & \multicolumn{6}{c}{Blur}                      & \multicolumn{6}{c}{Mask} \\
\cmidrule{2-21}          & \multicolumn{1}{c}{Dice} & \multicolumn{1}{c}{ASSD} & \multicolumn{3}{c}{Dice} & \multicolumn{3}{c}{ASSD} & \multicolumn{3}{c}{Dice} & \multicolumn{3}{c}{ASSD} & \multicolumn{3}{c}{Dice} & \multicolumn{3}{c}{ASSD} \\
\cmidrule{2-21}          & \multicolumn{1}{c}{0.00 } & \multicolumn{1}{c}{0.00 } & 0.50  & 1.00  & 1.50  & 0.50  & 1.00  & 1.50  & 3.00  & 5.00  & 7.00  & 3.00  & 5.00  & 7.00  & 0.10  & 0.25  & 0.40  & 0.10  & 0.25  & 0.40  \\
    \midrule
    U~\cite{Unet15}     & 83.78  & 2.52  & 81.46  & 70.39  & 61.53  & 2.46  & 3.47  & 5.37  & 75.78  & 68.16  & 61.53  & 4.23  & 7.10  & 9.79  & 81.00  & 70.98  & 55.88  & 2.89  & 4.09  & 5.60  \\
    V~\cite{VNet}     & 85.17  & 2.23  & 83.19  & 80.17  & 68.37  & 2.37  & 2.92  & 4.79  & 82.90  & 76.16  & 70.61  & 3.74  & 6.92  & 8.44  & 81.72  & 73.62  & 61.03  & 3.02  & 3.61  & 5.28  \\
    AU~\cite{AttentionU18}    & 84.71  & 2.44  & 82.24  & 74.73  & 63.31  & 2.46  & 5.34  & 6.87  & 82.48  & 76.94  & 73.46  & 4.55  & 6.71  & 7.26  & 81.83  & 73.93  & 59.39  & 2.77  & 3.23  & 5.49  \\
    nnU~\cite{isensee2021nnu}   & 85.10  & 3.03  & 62.30  & 44.93  & 39.08  & 13.95  & 22.18  & 26.60  & 82.11  & 74.67  & 70.13  & 5.00  & 7.80  & 8.90  & 81.68  & 73.31  & 60.91  & 3.12  & 4.03  & 5.46  \\
    nnFormer~\cite{zhou2023nnformer} & 85.79  & 1.89  & 84.09  & 77.75  & 69.62  & 2.14  & 3.92  & 4.53  & 83.96  & 81.81  & 77.26  & 2.32  & 3.32  & 4.88  & 80.59  & 67.97  & 51.51  & 2.26  & 3.12  & 5.71  \\
    U+TBraTS~\cite{zou2022tbrats} & 84.84  & 2.62  & 84.51  & 78.76  & 69.84  & 3.91  & 4.76  & 5.49  & 83.99  & 79.52  & 74.22  & 4.82  & 6.77  & 8.60  & 80.68  & 70.95  & 57.77  & 3.71  & 4.65  & 5.19  \\
    U+Our & 85.00  & 2.46  & 84.97  & 79.88  & 71.82  & 3.05  & 4.39  & 5.32  & 84.04  & 81.82  & 78.56  & 3.62  & 4.21  & 4.95  & 82.37  & 72.72  & 58.23  & 2.77  & 3.43  & 5.06  \\
    V+Our & 87.09  & \textbf{1.58}  & 83.80  & 81.97  & 72.11  & 2.06  & 2.47  & 4.10  & 82.75  & 77.57  & 73.83  & 4.29  & 6.70  & 7.57  & 83.71  & \textbf{75.69}  & \textbf{63.59}  & 2.10  & 2.70  & 4.30  \\
    AU+Our & 85.88  & 1.89  & 85.05  & \textbf{84.02}  & \textbf{80.34}  & \textbf{1.76}  & \textbf{1.84}  & \textbf{2.25}  & 85.45  & 83.02  & 79.15  & 2.51  & 3.71  & 4.83  & 81.44  & 74.38  & 61.79  & 2.35  & 3.11  & 5.42  \\
    nnU+Our & 85.51  & 3.03  & \textbf{85.37}  & 73.63  & 63.58  & 2.48  & 8.11  & 11.48  & 83.63  & 79.79  & 74.57  & 3.12  & 4.35  & 6.34  & 81.85  & 71.31  & 61.53  & 2.57  & 3.27  & 4.66  \\
    nnFormer+Our & \textbf{87.33}  & 1.69  & 84.92  & 81.18  & 71.16  & \textbf{1.76}  & 3.60  & 4.27  & \textbf{86.14}  & \textbf{83.44}  & \textbf{79.83}  & \textbf{1.91}  & \textbf{2.87}  & \textbf{3.38}  & \textbf{83.90}  & 74.12  & 58.78  & \textbf{1.88}  & \textbf{2.63}  & \textbf{3.76}  \\
    \bottomrule
    \end{tabular}%
    }
  \label{tab_brats1}%
\end{table*}%

Finally, we compared our proposed method with other uncertainty estimation methods on this dataset. As shown in Tab.~\ref{tab_brats2}, based on the ECE and UEO metrics, the performance of uncertainty-based methods gradually deteriorates with increasing levels of Gaussian noise, Gaussian blur, and random masking. However, methods integrated with DEviS exhibit slower degradation, especially for UNet, Attention-UNet, and nnFormer. Fig.~\ref{F_livsbrats} \textbf{b} provides a more intuitive comparison of the uncertainty estimation results for brain tumors among different methods. It is observed that the integrated DEviS-based network architecture provides more reliable uncertainty estimation, manifested in robust segmentation results under both normal and degraded conditions. Additionally, it offers corresponding uncertainty estimations and provides more pronounced uncertainty estimations for pixels affected by noise or masking. 

\begin{table*}[htbp]
  \centering
  \caption{The quantitative comparisons with uncertainty estimation based methods on the BraTS2019 dataset under differing Gaussian noise: ECE and UEO metrics with different Gaussian noise $\sigma=\left\{ {0,0.5,1.0,1.5} \right\}$, Gaussian blur ($\left\{ {\left( {{\sigma},k} \right)} \right\} = \left\{ {\left( {{\rm{3, 3}}} \right),  {\left( {{\rm{5, 5}}} \right)} ,\left( {{\rm{7, 7}}} \right)} \right\}$ and different mask ratio ($MR=\left\{ {0.1, 0.25, 0.4} \right\}$)). The numbers in the table need to be multiplied by ($0.01 \times $).}
    \resizebox{0.95\textwidth}{!}{
  \begin{tabular}{lrrrrrrrrrrrrrrrrrrrr}
    \toprule
    \multicolumn{1}{c}{\multirow{3}[6]{*}{Method}} & \multicolumn{2}{c}{Normal} & \multicolumn{6}{c}{Noise}                     & \multicolumn{6}{c}{Blur}                      & \multicolumn{6}{c}{Mask} \\
\cmidrule{2-21}          & \multicolumn{1}{c}{ECE} & \multicolumn{1}{c}{UEO} & \multicolumn{3}{c}{ECE} & \multicolumn{3}{c}{UEO} & \multicolumn{3}{c}{ECE} & \multicolumn{3}{c}{UEO} & \multicolumn{3}{c}{ECE} & \multicolumn{3}{c}{UEO} \\
\cmidrule{2-21}          & \multicolumn{1}{c}{0} & \multicolumn{1}{c}{0} & 0.5   & 1     & 1.5   & 0.5   & 1     & 1.5   & 3     & 5     & 7     & 3     & 5     & 7     & 0.1   & 0.25  & 0.4   & 0.1   & 0.25  & 0.4 \\
    \midrule
    PU~\cite{2018probabilisticU}    & \textbf{0.56}  & 87.43  & 0.85  & 1.06  & 1.43  & 82.93  & 78.54  & 62.20  & 0.64  & 0.81  & 1.01  & 85.94  & 81.41  & 77.11  & \textbf{0.69}  & 1.15  & 1.45  & \textbf{85.52}  & 76.67  & 64.97  \\
    UE~\cite{ensemble17}    & 0.82  & 85.73  & 1.47  & 1.83  & 2.11  & 82.68  & 77.07  & 72.70  & 1.45  & 1.95  & 2.37  & 82.69  & 77.06  & 73.67  & 1.02  & 1.45  & 1.87  & 83.24  & 74.91  & 63.10  \\
    DU~\cite{17dropoutCV}    & 0.72  & 85.58  & 0.86  & 0.95  & 1.14  & 84.96  & 78.03  & 69.68  & 0.79  & 0.86  & \textbf{0.93}  & 83.84  & 80.23  & 77.69  & 1.15  & 1.50  & 2.31  & 82.20  & 73.08  & 59.22  \\
    BQNAT~\cite{roy2019bqnat} & 0.62  & 84.94  & 0.64  & 0.79  & 1.01  & 84.44  & 79.94  & 70.13  & 0.64  & 0.81  & 0.91  & 83.84  & 81.72  & 78.64  & 0.70  & \textbf{0.93}  & 1.25  & 82.88  & 75.39  & 61.62  \\
    TTA~\cite{wang2018TTA}   & 0.64  & 84.84  & 0.62  & 0.86  & 1.03  & 85.95  & 77.10  & 71.65  & 0.68  & 0.84  & 0.95  & 83.36  & 80.52  & 77.75  & 0.73  & 1.01  & 1.32  & 81.43  & 73.35  & 60.35  \\
    U+Our & 0.58  & 86.67  & 0.59  & 0.68  & 0.88  & 86.27  & 84.48  & \textbf{81.82}  & 0.63  & 0.73  & 0.84  & 85.04  & 82.55  & 79.73  & 0.72  & 0.99  & 1.30  & 83.13  & 78.00  & 68.93  \\
    V+Our & 0.58  & 86.52  & 0.71  & 0.89  & 1.27  & 85.80  & 83.97  & 76.12  & 0.72  & 1.02  & 1.35  & 82.95  & 78.68  & 74.28  & \textbf{0.69}  & \textbf{0.93}  & \textbf{1.22}  & 83.44  & 78.59  & 69.03  \\
    AU+Our & 0.61  & 86.40  & 0.72  & 0.78  &\textbf{0.93}  & 84.71  & 83.73  & 81.78  & 0.67  & 0.76  & 0.85  & 85.34  & 83.14  & 79.27  & 0.74  & 1.03  & 1.36  & 83.15  & 77.26  & 67.57  \\
    nnU+Our & 0.58  & 86.67  & 0.61  & 1.17  & 1.85  & 86.36  & 73.71  & 63.60  & 0.68  & 0.88  & 1.14  & 84.28  & 80.11  & 74.71  & 0.70  & 0.94  & 1.25  & 83.16  & 75.31  & 64.26  \\
    nnFormer+Our & \textbf{0.56}  & \textbf{88.19}  & \textbf{0.58}  & \textbf{0.70}  & 0.94  & \textbf{87.81}  & \textbf{86.51}  & 81.59  & \textbf{0.59}  & \textbf{0.69}  & \textbf{0.83}  & \textbf{87.15}  & \textbf{83.76}  & \textbf{80.15}  & \textbf{0.69}  & 0.94  & 1.28  & 85.39  & \textbf{79.72}  & \textbf{72.40}  \\
    \bottomrule
    \end{tabular}%
    }
  \label{tab_brats2}%
\end{table*}%

\subsection{DEviS ensures computationally efficient uncertainty estimation}
\begin{figure}
\centering
\includegraphics[width=1\linewidth]{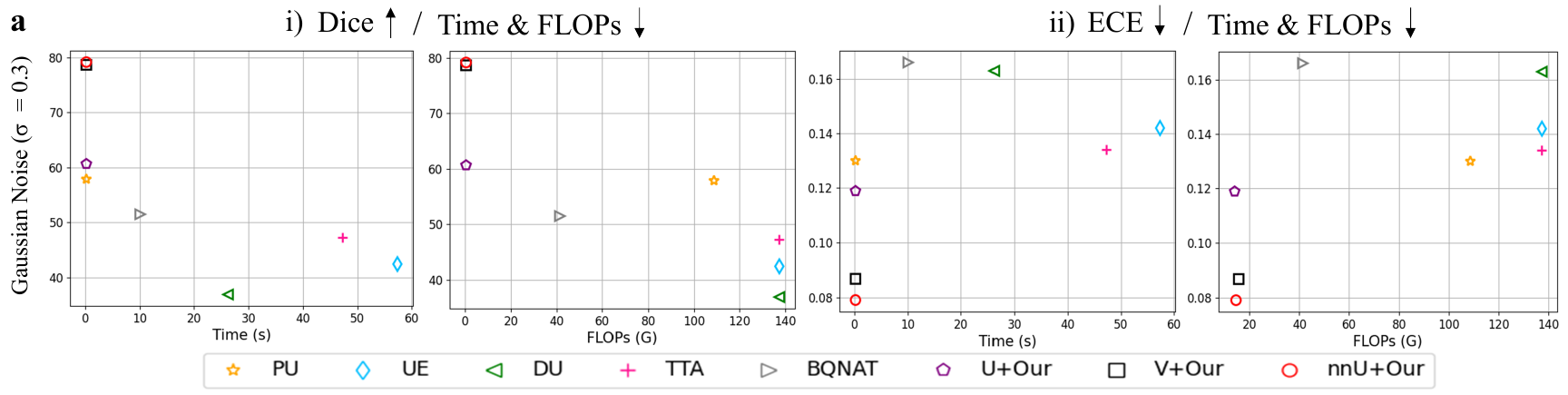}
\includegraphics[width=1\linewidth]{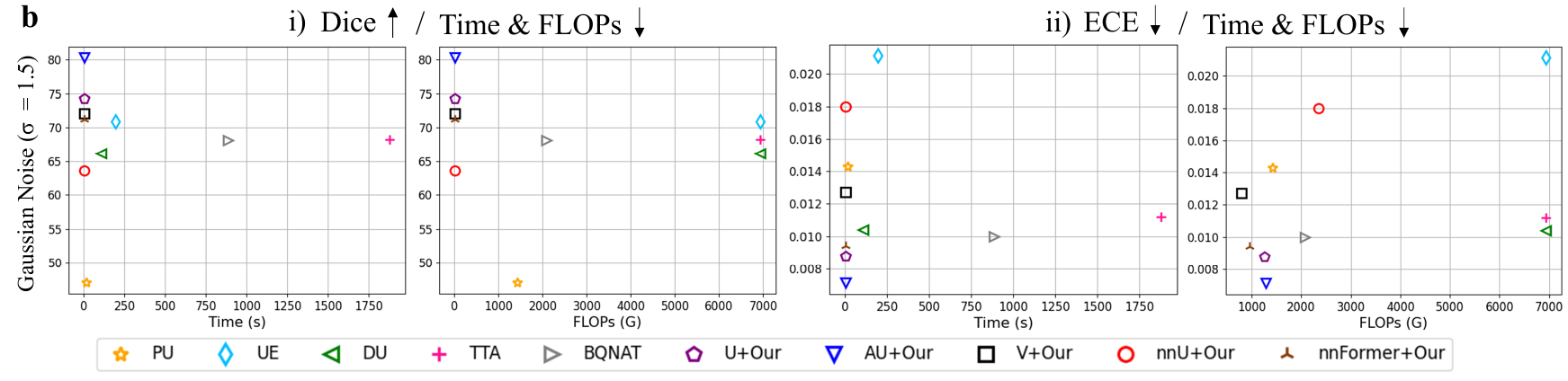}
\caption{Inference analysis of Uncertainty estimation models on the ISIC2018 and BraTS2019 datasets. \textbf{a.} Inference analysis of Uncertainty estimation models on the ISIC2018 dataset under Gaussian noise condition (${\sigma}{\rm{ = }}0.3$). \textbf{b.} Inference analysis of Uncertainty estimation models on the BraTS2019 dataset under Gaussian noise condition (${\sigma}{\rm{ = }}1.5$). }
\label{F_time}
\end{figure}

To demonstrate the effectiveness of the efficiency, we provide more insight into the computational cost performances of well-known uncertainty estimation methods (DU~\cite{17dropoutCV}, UE~\cite{ensemble17}, PU~\cite{2018probabilisticU}, TTA~\cite{wang2018TTA} and BQNAT~\cite{roy2019bqnat}) on the datasets of ISIC2018, and BraTS2019 with Gaussian noise by ${\sigma}{\rm{ = }} \left\{{0.3,0.2,1.5} \right\}$ (Fig.~\ref{F_time} (a) and Fig.~\ref{F_time} (b)). Among the evaluated methods, UE and TTA are the least computationally efficient, yet they yield slightly better segmentation results compared to DU. BQNAT demonstrates slightly higher calculation efficiency than DU, with comparable segmentation performance to TTA. Although PU has shown improvements in both computing efficiency and segmentation results, it still requires additional sampling time for testing. After applying DEviS, fundamental deep learning models such as U-Net, Attention-UNet, V-Net, nnUNet, and Transformer segmentation methods represented by nnFormer exhibit noticeable improvements in both testing time and FLOPs compared to the previously mentioned uncertainty quantification methods. Additionally, they deliver more robust segmentation outcomes post-integration. Comprehensive numerical analyses of computational efficiency, segmentation performance, and uncertainty estimation reliability, including comparisons across multiple models and metrics, are provided in the Supplementary Materials.

\subsection{Extending DEviS to Semi-Supervised Medical Image Segmentation}
\begin{figure}
\centering
\includegraphics[width=1\linewidth]{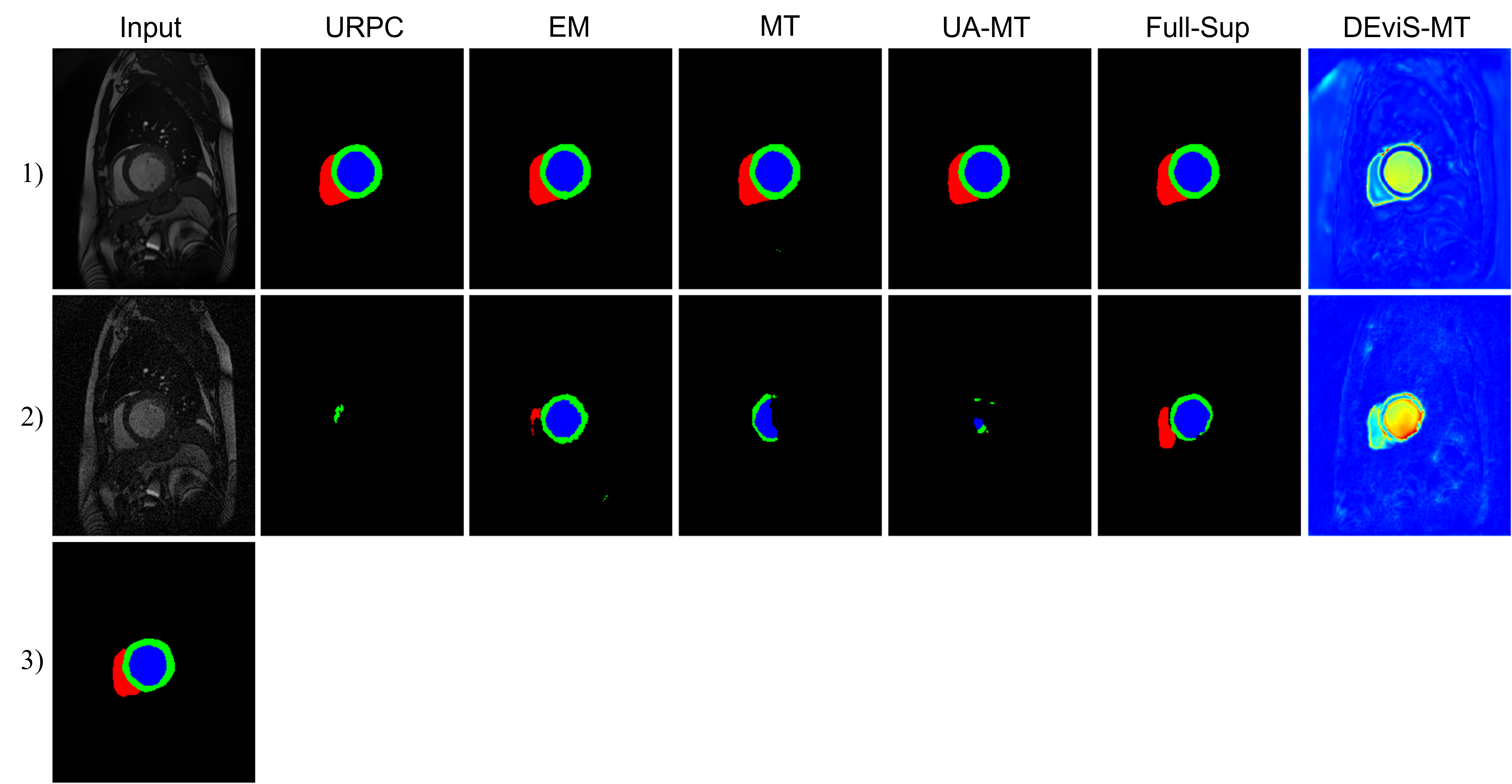}
\caption{Visual comparisons of the proposed method with typical approaches on the ACDC2018 dataset under a semi-supervised setting.}
\label{F_acdc}
\end{figure}

Subsequently, we explored the potential of adapting DEviS to semi-supervised medical image segmentation. Following the protocol of~\cite{luo2022semi}, we conducted experiments on the ACDC2018 dataset using 10\% of the labeled data. To demonstrate the generalizability of DEviS, U-Net was chosen as the backbone model for these experiments. The results are presented in Tab.~\ref{tab_semi}. We compared our approach with several representative methods, including EM, MT, UA-MT and Full supervised method (Full-Sup). To provide a clearer comparison between different methods, our experiments mainly report results for the RV cavity. As shown in Table\ref{tab_semi}, DEviS achieves performance comparable to MT under clean conditions, and surpasses UA-MT in terms of Dice coefficient. Under noisy conditions, where Gaussian noise with a variance of 0.05 was added to the input images, the DEviS-equipped MT model demonstrated significantly enhanced robustness, outperforming UA-MT and yielding Dice improvements of 6.52\% and 17.56\% over the MT and Full-Supervised baselines, respectively. Furthermore, visual comparisons in Fig.~\ref{F_acdc} reveal consistent findings. Under standard settings, DEviS yields segmentation results comparable to other methods, while under noisy conditions ($\sigma$ = 0.05), it demonstrates superior performance. These results indicate that DEviS is well-suited for semi-supervised medical image segmentation and provides enhanced robustness under input perturbations.

\begin{table}[htbp]
  \centering
  \caption{Comparison of the proposed method with typical approaches on the ACDC2018 dataset under a semi-supervised setting.}
    \begin{tabular}{ccccccc}
    \toprule
    \multirow{2}[2]{*}{Method} & \multicolumn{3}{c}{Normal} & \multicolumn{3}{c}{Noise (0.05)} \\
          & Dice  & ASSD  & HD95  & Dice  & ASSD  & HD95 \\
    \midrule
    EM~\cite{vu2019advent} & 79.20  & 1.36  & 5.01  & 44.70  & 8.88  & 17.16 \\
    URPC~\cite{luo2022semi}  & 78.35  & \textbf{0.74} & \textbf{3.15} & 21.51  & 20.35 & 31.32  \\
    MT~\cite{tarvainen2017mean} & \textbf{80.76} & 1.92 & 7.81  & 28.53  & 18.02  & 22.24 \\
    UAMT~\cite{yu2019uncertainty} & 70.93  & 0.87  & 4.69  & 26.88  & 19.16  & 26.15 \\
    DEviS-MT & 80.25  & 1.29  & 7.03  & \textbf{51.22} & \textbf{7.94} & \textbf{13.13} \\
    \midrule
    Full-Sup & 90.70  & 0.35  & 3.42  & 33.66  & 16.70  & 22.42 \\
    \bottomrule
    \end{tabular}%
  \label{tab_semi}%
\end{table}%

\subsection{DEviS empowers clinical application}
\textbf{1) Out-of-distribution detector:} In this section, the proposed DEviS method is first demonstrated to be applicable for OOD detection. It is essential that image processing systems identify any OOD samples in clinical settings. Uncertainty estimation quantifies the uncertainty of the in-distribution (ID) and OOD data to detect inputs that are far outside the training data distribution. DEviS can thus be used to alert clinicians to areas where lesions may be present in OOD data. Further, we equipped DEviS with UAF to distinguish ID and OOD samples. 

\begin{figure}
\centering
\includegraphics[width=1\linewidth]{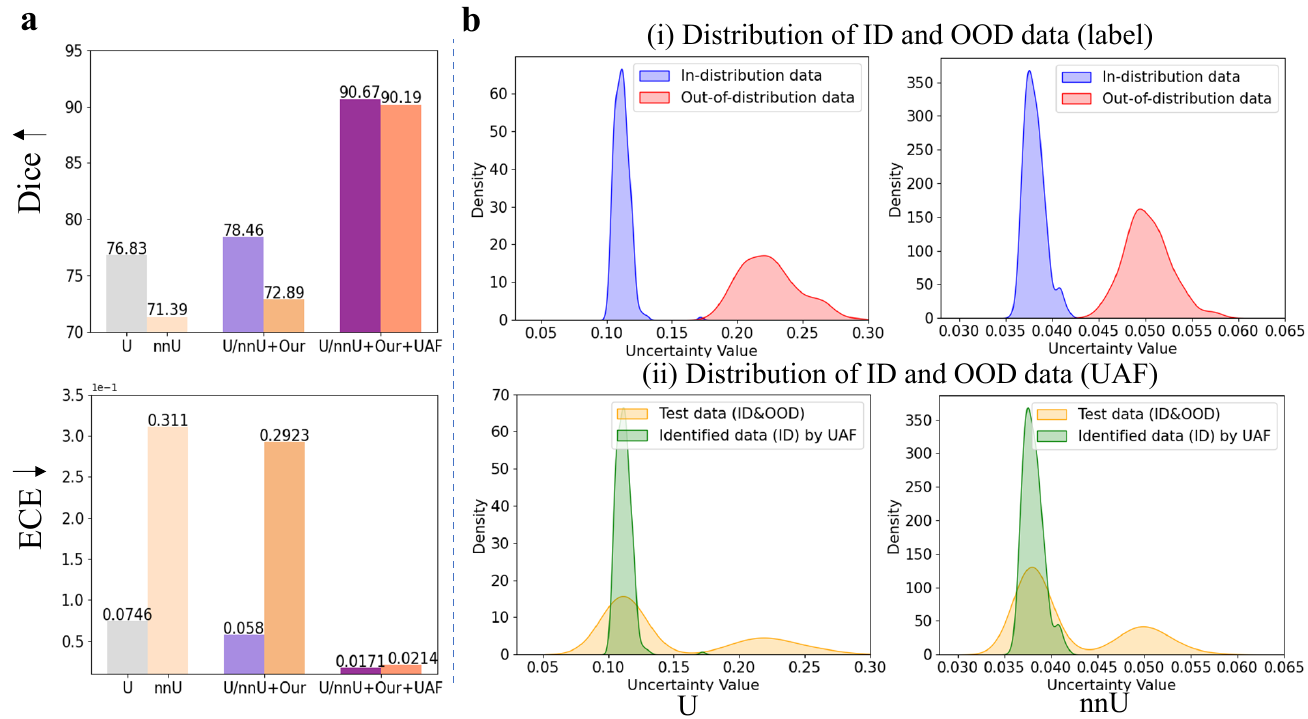}
\includegraphics[width=1\linewidth]{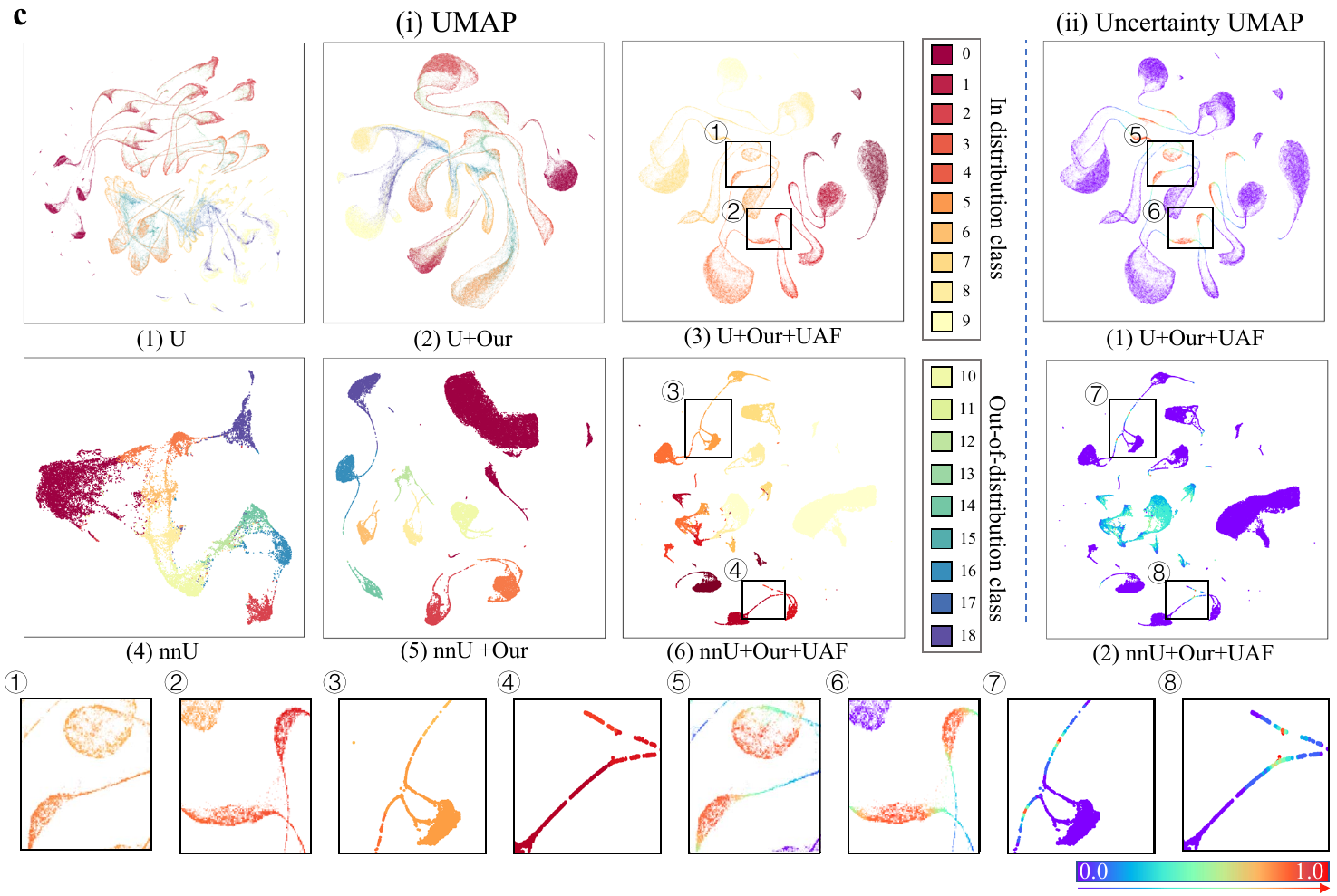}
\caption{Clinical application for OOD detector. \textbf{a.} The qualitative results of Dice (\%) and ECE on different qualities fundus images. \textbf{b.} Density of uncertainty of predictions for ID and OOD data. \textbf{c.} The UMAP plots of latent representations before/after integrating our method. }
\label{F_5}
\end{figure}

We conducted OOD experiments on the Johns Hopkins OCT dataset and Duke OCT dataset with DME. Fig.~\ref{F_5} illustrates the clinical validation of OOD detection. Specifically, Figure 5 \textbf{a} compares the segmentation performance of U-Net and nnU-Net, both before and after applying our method with the UAF strategy, using Dice and ECE metrics. Fig.~\ref{F_5} \textbf{b} highlights the differences in predicted ID and OOD distributions relative to the ground truth labels under different conditions: Fig.~\ref{F_5} \textbf{b (i)} shows the ID and OOD distributions predicted by U-Net and nnU-Net based on the ground truth labels, while Figure Fig.~\ref{F_5} \textbf{b}(ii) compares the data distributions before and after applying the UAF strategy. Lastly, Fig.~\ref{F_5} \textbf{c} uses UMAP to visualize the differences between segmentation classes, where Fig.~\ref{F_5} \textbf{c (i)} depicts the standard UMAP visualization, and Fig.~\ref{F_5} \textbf{c (ii)} provides an uncertainty-aware version of the UMAP visualization. As shown in Fig.~\ref{F_5} \textbf{a}, we first observed a slight improvement in results for mixed ID and OOD data after using DEviS. Then, we found significant differences in the performance of the segmentation between the with or without UAF. Additionally, there were also marked differences in the distribution of uncertainty between the ID and OOD data, especially adding the UAF module as shown in Fig.~\ref{F_5} \textbf{b}. As depicted in Fig.~\ref{F_5} \textbf{c (i)}, We then employed Uniform Manifold Approximation and Projection (UMAP) to visually assess the integration of our method. In the spatial clustering results of the base network, we observed overlapping of ID and OOD data batches. After integrating DEviS, we observed improved batch-specific separation of ID and OOD data, particularly for the ID data. Furthermore, the integration of UAF with DEviS effectively eliminated the OOD, resulting in a more pronounced batch effect. Additionally, we first presented the uncertainty estimation map corresponding to UMAP in the Fig.~\ref{F_5} \textbf{c (ii)}. It is evident from the map that the boundary region between different batches exhibits significantly higher uncertainty. 

\textbf{2) Image quality indicator:} Then, DEviS can serve as an indicator for representing the quality of medical images. Uncertainty estimation is an intuitive and quantitative way to inform clinicians or researchers about the quality of medical images. DEviS guides image quality quantitatively through the distribution of uncertainty values and qualitatively through the degree of explicitness of the uncertainty map. Furthermore, our developed UAF module aids in the initial screening of low-quality and high-quality data. High-quality data can be directly employed in clinical practice, while low-quality data necessitates expert judgment before utilization.

\begin{figure}
\centering
\includegraphics[width=1\linewidth]{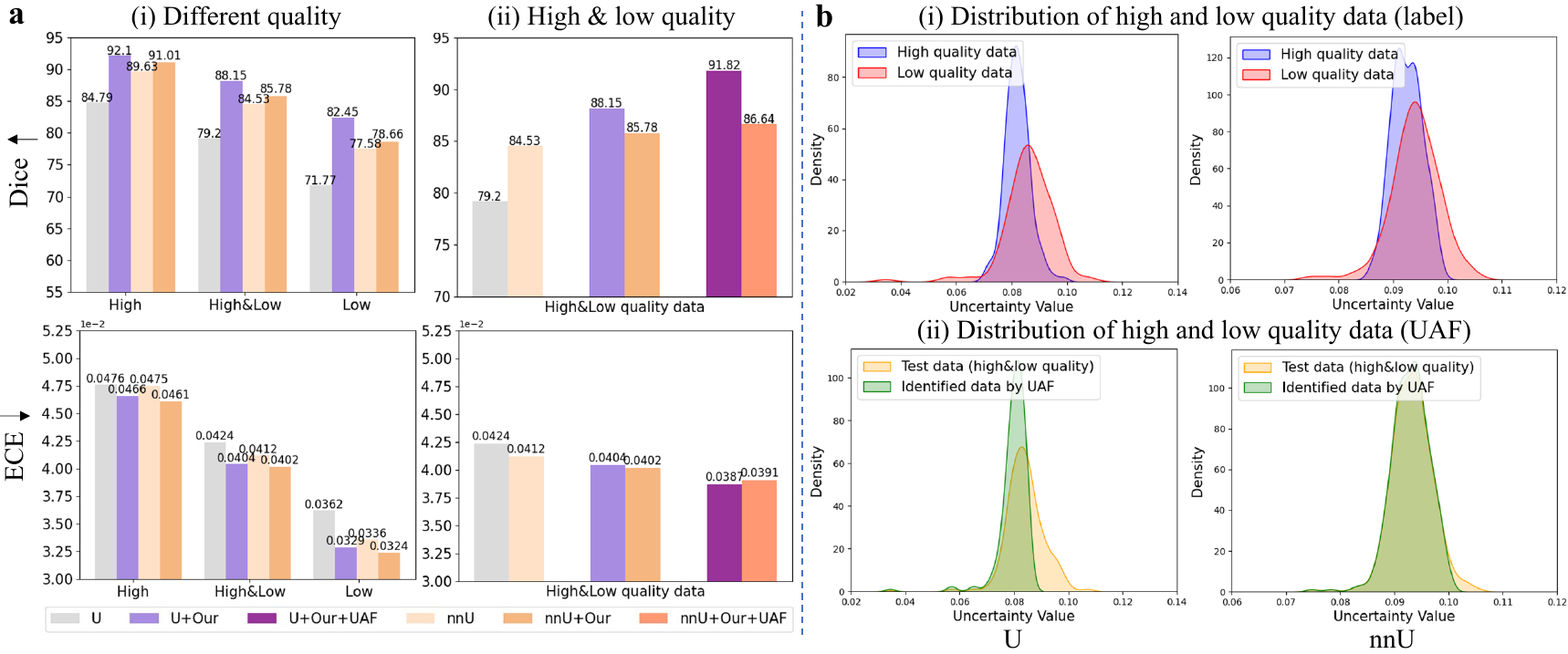}
\includegraphics[width=1\linewidth]{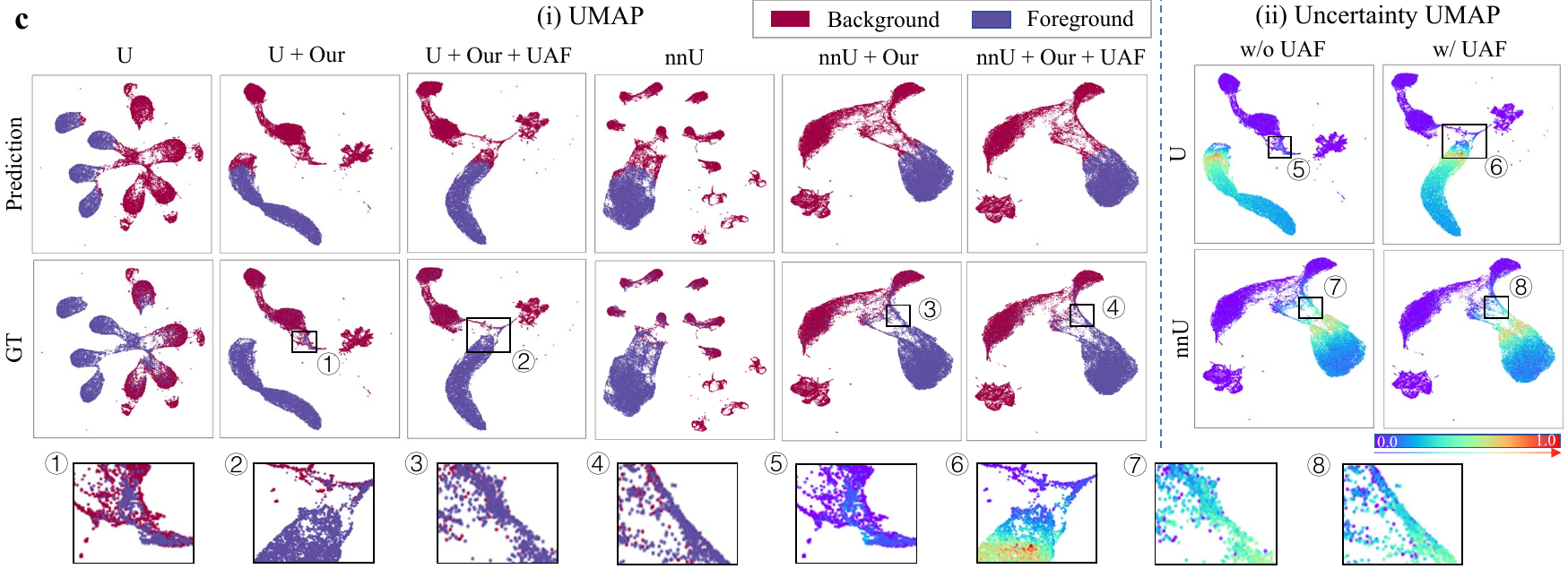}
\caption{Clinical applications for quality indicator. \textbf{a.} The qualitative results of Dice (\%) and ECE on different qualities fundus images. \textbf{b.} Density of uncertainty of predictions for different quality data. \textbf{c.} UMAP differences between different methods. }
\label{F_6}
\end{figure}

In what follows, we apply DEviS with UAF to indicate the quality of data for real-world applications. The FIVES datasets are used for quality assessment experiments. We initially classified samples into three categories based on their quality labels: high quality, high \& low quality, and low quality. We observed distinct performance variations among these categories (Fig.~\ref{F_6} \textbf{a} (i)). To further demonstrate its ability to indicate image quality, we delved into a combination of high and low-quality data to filter out high-quality data. Before the application of UAF, we identified 159 high-quality and 141 low-quality data samples. Upon implementing UAF, the distribution shifted, resulting in 153 high-quality and 61 low-quality data samples. This transition led to a remarkable increase in the proportion of high-quality data from 53\% to 71\%. Notably, the task at hand posed a greater challenge in assessing data quality compared to the detection of OOD data, as all data sources originated from the same distribution. We also found a significant performance boost with UAF in Dice and ECE metrics. (Fig.~\ref{F_6} \textbf{a} (ii)). Additionally, we investigated the distribution of uncertainty to discern differences between different qualities data (Fig.~\ref{F_6} \textbf{b} (i)). Moreover, the uncertainty distribution of high and low mixed quality with UAF was closer to the low-quality data (Fig.~\ref{F_6} \textbf{b} (ii)). The spatial clustering results of mixed-quality images were visualized using UMAP in the Fig.~\ref{F_6} \textbf{c}. Prior to incorporating our algorithm, some batch-specific separation was observed, albeit with partially overlapping regions (Fig.~\ref{F_6} \textbf{c} (i) 1st and 4th columns). However, upon integrating DEviS with UAF, a slight batch effect was observed (Fig.~\ref{F_6} \textbf{c} (i) 2nd, 3rd, 5th and 6th columns). Additionally, the UMAP visualization with uncertainty map exhibited uncertainty warnings for partially overlapping points, with noticeably high uncertainties along the edges of prediction errors (Fig.~\ref{F_6} \textbf{c} (ii) (1, 2)). 

More intuitive segmentation results and uncertainty maps for both ID and OOD data, as well as for low- and high-quality images, are provided in the Supplementary Materials. These results further demonstrate that DEviS with UAF can filter abnormal lesion-prone regions in OOD data and serve as an image quality indicator for fair valuation of personal data.

\section{Conclusion}
In this study, we conducted a comprehensive analysis of the foundational DEviS model, establishing a robust framework for reliable medical image segmentation. Our approach enhances the identification of trustworthy segmentation regions, ultimately fostering confidence between healthcare professionals and deep learning models. Our framework is adaptable to different network architectures, including U-Net-based, Transformer-based, and Mamba-based models. It achieves accurate and well-calibrated segmentation results while maintaining high computational efficiency in uncertainty estimation. We first model uncertainty for different segmentation classes using Subjective Logic theory. Additionally, we developed the CUP module to improve confidence calibration while preserving the segmentation performance of the base network. Furthermore, the proposed UAF strategy offers a reliable mechanism to ensure the model’s effective integration into clinical applications, such as OOD data detection and image quality assessment. We rigorously evaluated DEviS across multiple publicly available datasets, assessing its accuracy, robustness, and reliability. In addition, we extended DEviS to semi-supervised segmentation settings, where it demonstrated strong robustness under noisy conditions. Our results demonstrate that DEviS, combined with UAF, achieves superior performance with enhanced interpretability, providing reliable diagnostic support for OOD data. We believe that our work lays a valuable foundation for future research in trustworthy medical AI.

While our proposed foundational module has demonstrated strong calibration and robustness across diverse segmentation settings (2D, 3D, and multi-modal), including semi-supervised scenarios and OOD detection, its applicability to highly heterogeneous clinical edge cases, as represented in RAOS~\cite{luo2024rethinking}, remains an important direction for future validation. In subsequent work, we plan to integrate and assess our method on RAOS to further explore its robustness and generalization capabilities under real-world complexities.

\bibliographystyle{IEEEtran}
\bibliography{uncertainty.bib}

\newpage
\section*{Supplementary Materials}
\setcounter{figure}{0}
\setcounter{table}{0}
\setcounter{equation}{0}
\setcounter{page}{1}
\renewcommand{\thetable}{S\arabic{table}}
\renewcommand{\thefigure}{S\arabic{figure}}
\renewcommand{\theequation}{S\arabic{equation}}

 \subsection{Algorithm 1: Training Procedure of DEviS}
The details of the training process of the proposed DEviS method are presented in Algorithm~\ref{alg:DEviS}.
\begin{algorithm}[ht]
    \caption{~The training process of proposed DEviS model}
    \label{alg:DEviS}
    \begin{algorithmic}[1]
        \STATE \textbf{Given:} a medical image domain training dataset $\mathbb{X}=\left\{\{{x}_m\}_{m=1}^M, y_x\right\}$, initialized medical segmentation backbone ${f}\left(  \cdot  \right)$, hyperparameter ${\lambda}=0.02$, and epochs for training $T$.	
        \FOR{$t=1,\ldots,T$}
            \FOR{$m=1,\ldots,M$}
                \STATE Perform deep evidential feature extraction (Eq.~\ref{E_0} \& Eq.~\ref{E_1});
                \STATE Characterize Dirichlet distribution (Eq.~\ref{E_5});
                \STATE Estimate evidential uncertainty (Eq.~\ref{E_6});
                \STATE Compute evidential deep learning loss (Eq.~\ref{E_16});
                \STATE Compute soft Dice loss loss;
                \STATE Compute calibration uncertainty loss (Eq.~\ref{E_19});
                \STATE Compute total evidence calibration training loss (Eq.~\ref{E_20});
                \STATE Update gradients of the base segmentation network;
            \ENDFOR
        \ENDFOR
        \STATE \textbf{Return:} Network parameters
    \end{algorithmic}
\end{algorithm}

\subsection{Experimental datasets}
1) ISIC2018 dataset. First, the public available International Skin Imaging Collaboration (ISIC) 2018 dataset~\cite{ISIC2018} is validated for reliable 2D medical image segmentation task. Following the~\cite{2020CA}, a total of 2594 images and their ground truth are randomly divided into a training set, validation set, and test set, containing 1814, 260 and 520 images, respectively. To verify the robustness and credibility of different models under OOD conditions, we add different levels of Gaussian noise and random masks to the test set, and perform 5-fold cross-validation for final results. First, we added the standard deviation of the Gaussian noise ranging from [0.1, 0.2, 0.3, 0.4, 0.5] to the original data with normalization.

2) KiTS2021~\cite{kits21} dataset. The Kidney Tumor Segmentation 2021 (KiTS21) dataset is a publicly available benchmark for automated kidney and kidney tumor segmentation in contrast-enhanced 3D abdominal CT scans. It consists of 300 cases, which we split into training, validation, and test sets in a 7:1:2 ratio. Each case was further divided into 3D patches of size $16 \times 256 \times 256$. The choice of 16 was made due to the memory limitations of computational resources. To validate the robustness of the proposed method, we also constructed simulated experimental data with Gaussian noise and Gaussian blur. Specifically, Gaussian noise with a standard deviation of 0.1 was added to the normalized data. Additionally, Gaussian blur with a variance of 7 and a kernel size of $(7,7)$ was applied to the test data.

3) LiTS2017 dataset. The Liver Tumor Segmentation (LiTS) Challenge 2017 serves as a benchmark for reliable 3D medical image segmentation. It consists of 131 publicly available contrast-enhanced 3D abdominal CT scans for training and 70 for testing. Following~\cite{HDenseU}, all samples were resampled to a uniform resolution of $16 \times 256 \times 256$ with a voxel spacing of $0.69 \times 0.69 \times 1$. The dataset was randomly split into a training set and a test set, containing 105 cases (approximately 985 volumes) and 26 cases (approximately 245 volumes), respectively. To assess robustness, we constructed a simulated dataset by introducing varying levels of Gaussian noise, Gaussian blur, and random masking into the test data of the 3D volumes. Specifically, Gaussian noise was added to the normalized data with a standard deviation of 0.4. Gaussian blur was applied to the test data with a variance of 15 and a kernel size of 20, denoted as $(15, 20)$. Additionally, a random masking strategy was implemented, utilizing an 8-pixel mask size and a mask ratio of 0.1 on the original data.

4) BraTS2019 dataset. More importantly, the Brain Tumor Segmentation (BraTS) 2019 challenge~\cite{BraTSbench} with varying degradation conditions(such as noise, blur and mask) are constructed for reliable multi-modality 3D medical image segmentation tasks. Four modalities of brain MRI scans with a volume of $240 \times 240 \times 155$ are used. 335 cases of patients on BraTS2019 with ground-truth are randomly divided into train dataset, validation dataset, and test dataset with nearly 265, 35, and 35 cases, respectively.  The three tumor sub-compartment labels are combined to segment the whole tumor and all inputs are uniformly adjusted to $128\times128\times128$ voxels during the training.  The outputs of our network contain 4 classes, which are background (label 0), necrotic and non-enhancing tumor (label 1), peritumoral edema (label 2), and GD-enhancing tumor (label 4). Similarly, in order to verify the reliability uncertainty estimation and robust segmentation results of the model under OOD data, three changes were made to the test set, namely Gaussian noise, Gaussian blur, and random mask. Gaussian noise is added to the normalized data with standard deviation of the ranging from [0.5, 1.0, 1.5, 2.0]. Gaussian blur is added to the test data with variance varying from 3 to 9 and kernel sizes of 3 to 9, specifically ranging from [(3, 3), (5, 5), (7, 7), (9, 9)]. The strategy of random mask with 8 pixel-size ranging from [0.1, 0.25, 0.4, 0.6] are also deployed for the original data.

5) Johns Hopkins OCT \& Duke-OCT-DME datasets. Johns Hopkins OCT (JH-OCT) dataset (\url{https://iacl.ece.jhu.edu/index.php?title=Main_Page}) is the healthy controls and multiple sclerosis dataset, provided by Johns Hopkins University, comprises OCT scans from 35 subjects, including both healthy individuals and patients with multiple sclerosis. Each OCT scan is annotated with eight distinct retinal layers, providing detailed structural information critical for understanding retinal changes associated with multiple sclerosis. This dataset is particularly valuable for advancing segmentation algorithms and studying the impact of MS on retinal morphology, with potential applications in both research and clinical practice.\\
The Duke OCT dataset with Diabetic Macular Edema (Duke-OCT-DME) (\url{https://people.duke.edu/~sf59/Chiu_BOE_2014_dataset.htm}) consists of 110 annotated SD-OCT B-scan images, each measuring $512 \times 740$ pixels. These images are derived from 10 patients, with 11 B-scans captured per patient. For each patient, the B-scans are centered around the fovea, with additional frames obtained at five lateral positions on either side of the fovea. The dataset includes detailed annotations of retinal layers and fluid regions, carried out by two expert clinicians.

In clinical, the OOD sample and the value of data are essential for AI medicine. To facilitate clinical application, we utilized these two real-world datasets (JH-OCT dataset and Duke-OCT-DME dataset) to OOD experiments, aiming to validate the effectiveness of the proposed method. 35 cases of patients on JH-OCT with ground-truth are randomly divided into train dataset, validation dataset and test dataset with nearly 25, 5 and 5 cases, respectively. The 5 cases of test dataset on JH-OCT are used for ID detection. In particular, the 10 cases from the Duke-OCT-DME are used as another test dataset for OOD detection. Every case is uniformly adjusted to $128 \times 1024$ voxels during the training and testing.

6) FIVES dataset \& DRIVE dataset. In the FIVES dataset, each image is assessed with four quality evaluations: normal, illumination and color distortion, blur, and low contrast distortion. The dataset consists of 300 image slices, which include both high-quality and low-quality images. In the second application, the FIVES dataset is used as a quality indicator for retinal vessel segmentation. In this experiment, normal images are defined as high-quality data, while images under other conditions are classified as low-quality data. The performance of DEviS was then evaluated on a mixed dataset from FIVES, consisting of 159 high-quality image slices and 141 low-quality image slices.

The DRIVE (Digital Retinal Images for Vessel Extraction) dataset is widely used for retinal vessel segmentation tasks. It includes 40 high-resolution color fundus images of the retina, along with manual annotations for retinal vessel locations. In this study, the DRIVE dataset is employed to validate the generalization of the CUP module.

7) ACDC2018 dataset. The ACDC~\cite{bernard2018deep} dataset is a four-class medical image segmentation dataset comprising cardiac MRI scans from 100 patients. The segmentation labels are stored in NIfTI format and include background (label 0), right ventricle (RV) cavity (label 1), myocardium (label 2), and left ventricle (LV) cavity (label 3). The dataset is divided into fixed subsets of 70, 10, and 20 patient scans for training, validation, and testing, respectively. During training, we adopt U-Net as the backbone and use 2D input patches of size 256×256, with a zero-valued mask region of 170×170. Training parameters include a batch size of 24, 10,000 pre-training iterations, and 30,000 self-training iterations. To provide a clearer comparison among different methods, our experiments report results primarily on the RV cavity.

\subsection{Ablation study}\label{AS}
In this section, we first perform ablation studies on the parameters $\lambda$ and $W$, followed by a comparison between the calibration method $e^{-u}$ and the Softmax activation function. Finally, we evaluate the generalization of the CUP module and investigate the impact of different uncertainty thresholds in UAF on model performance.

1) Parameter Settings of ${\lambda _{{\rm{ini}}}}$ and $W$. For the parameters ${\lambda _{{\rm{ini}}}}$ and $W$, we performed ablation experiments using the representative V-Net and nnU-Net models on the ISIC2018 dataset, with Dice as the evaluation metric, as shown in Tables \ref{tab_lamda} and \ref{tab_w} . For ${\lambda _{{\rm{ini}}}}$, the results indicate that as its value increases, the model's segmentation performance initially improves but then declines, achieving the best performance at ${\lambda _{{\rm{ini}}}} = 0.02$. For $W$, we observe a clear downward trend in model performance as $W$ increases, with the optimal value found at $W = 1$, which aligns with findings from previous literature.

\begin{table}[htbp]
  \centering
  \caption{Ablation Study Results for ${\lambda _{{\rm{ini}}}}$ on the ISIC2018 dataset.}
    \begin{tabular}{lrrrrr}
    \toprule
    ${\lambda _{{\rm{ini}}}}$ & 0.01  & 0.02  & 0.1   & 0.2   & 0.5 \\
    \midrule
    V-Net & 84.45 & \textbf{86.57} & 84.26 & 82.28 & 83.05 \\
    nnU-Net & 86.7  & \textbf{87.31} & 87.17 & 87.06 & 86.98 \\
    \bottomrule
    \end{tabular}%
  \label{tab_lamda}%
\end{table}%

\begin{table}[htbp]
  \centering
  \caption{Ablation Study Results for $W$ on the ISIC2018 dataset.}
    \begin{tabular}{lrrrrr}
    \toprule
    $W$     & 1     & 5     & 10    & 50    & 100 \\
    \midrule
    V-Net & \textbf{86.57} & 84.7  & 83.53 & 83.41 & 83.24 \\
    nnU-Net & \textbf{87.31} & 87.28 & 87.14 & 86.51 & 86.47 \\
    \bottomrule
    \end{tabular}%
  \label{tab_w}%
\end{table}%

2) Comparison results with Calibrated method $e^{-u}$ and Softmax function.
To compare with other calibration methods, this paper evaluates the relatively simple $e^{-u}$ calibration technique. When incorporated into the training process, the $e^{-u}$ value decreases as $u$ reduces, thereby achieving uncertainty calibration. Additionally, we compared this method with the Softmax function. As seen in Tab.~\ref{tab_eu}, the $e^{-u}$ calibration method performs worse than the proposed CUP method across various metrics. Moreover, the Softmax function leads to an increase in ECE and other evaluation metrics, thereby reducing model performance, lowering model confidence, and resulting in overconfidence, which is consistent with the findings in~\cite{evidential18,2020trusted}.

\begin{table}[htbp]
  \centering
  \caption{Comparison of nnU-Net as the base model using the Calibrated method \(e^{-u}\) and the Softmax function on the ISIC2018 dataset.}
    \begin{tabular}{lrrrrr}
    \toprule
    Method      & Dice  &  ASSD  & ECE    & UEO \\
    \midrule
    $e^{-u}$ & 86.05 & 6.93  & 0.0538 & 0.9034\\
    Softmax & 87.31 & 87.28 & 87.14 & 86.51 \\
    Ours & \textbf{87.31} & 6.25 & 0.0469 & 0.9098 \\
    \bottomrule
    \end{tabular}%
  \label{tab:addlabel}%
\end{table}%

\begin{table}[htbp]
\centering
  \caption{Comparison of nnU-Net as the base model using the Calibrated method \(e^{-u}\) and the Softmax function on the ISIC2018 dataset.}
    \begin{tabular}{rcccccc}
    \toprule
    \multicolumn{1}{c}{\multirow{2}[4]{*}{Method}} & \multicolumn{3}{c}{Normal} & \multicolumn{3}{c}{Noise (0.3)} \\
\cmidrule{2-7}          & Dice  & ASSD  & ECE   & Dice  & ASSD  & ECE \\
    \midrule
    $e^{-u}$~\cite{tarvainen2017mean}     & 86.05 & 6.93  & 0.0538 & 73.68 & 14.2  & 0.0961 \\
    Softmax&  86.28   &   6.72 & 0.9103    &  70.37  & 13.63   & 0.1178 \\
    Ours& \textbf{87.31} & \textbf{6.25}  & \textbf{0.0469} & \textbf{78.88}  & \textbf{10.53}  & \textbf{0.0898}  \\
    \bottomrule
    \end{tabular}%
  \label{tab_eu}%
\end{table}%

3) Generalization of CUP. To assess the generalization capability of the proposed CUP strategy, we conducted experiments using the nnU-Net model on two retinal vessel segmentation datasets. Specifically, we trained the model on the FIVES dataset and directly tested it on the DRIVE dataset to examine its generalization performance. Then, segmentation accuracy using Dice and ASSD metrics were evaluted, while reliability of uncertainty estimation was assessed through ECE and UEO, as summarized in Table~\ref{tab_gene}. The results demonstrate that, compared to the standard nnU-Net, the proposed method enhances model generalization, and the CUP strategy offers further room for improvement. 
\begin{table}[htbp]
  \centering
  \caption{Generalization Performance of the proposed CUP Module on Retinal Vessel Segmentation. (Trained on the FIVES dataset and tested on the DRIVE dataset)}
    \begin{tabular}{lrrrrr}
    \toprule
    Method     & Dice     &  ASSD  & ECE    & UEO \\
    \midrule
    nnU-Net & 58.64 & 7.65 & 0.0654 & 0.3266 \\
    nnU-Net + TBraTS~\cite{zou2022tbrats} & 60.89 & 2.37 & 0.0639 &  0.3455 \\
    nnU-Net + TBraTS~\cite{zou2022tbrats} + CUP & \textbf{63.76} & \textbf{2.31} & \textbf{0.0620} & \textbf{0.3666}\\
    \bottomrule
    \end{tabular}%
  \label{tab_gene}%
\end{table}%

4) Comparison results of different uncertainty thresholds. Finally, we conducted ablation experiments to evaluate the effect of different UAF thresholds on OOD detection. The experimental results are presented in Table~\ref{tab_umax}, which compares the outcomes for the JHU dataset and another dataset. Specifically, we investigated the influence of various UAF thresholds, including ${0.9\bf{u}}_m^*$, $0.95{\bf{u}}_m^*$, and ${\bf{u}}_m^*$. It is important to note that since the uncertainty values range from 0 to 1, they exhibit significant variation within this range. As shown in the table, the optimal OOD performance is achieved when the threshold is set to U, while changing the threshold from $0.95{\bf{u}}_m^*$ to $0.9{\bf{u}}_m^*$ has minimal impact on OOD detection performance.

\begin{table}[htbp]
  \centering
  \caption{Comparison of different UAF thresholds on the ISIC2018 dataset using nnU-Net.}
    \begin{tabular}{lrrrrr}
    \toprule
    Thresholds  & Dice     &  ASSD  & ECE    & UEO \\
    \midrule
    0.9${\bf{u}}_m^*$ & 90.09 & 0.3574 & 0.0214 &  0.9962 \\
    0.95${\bf{u}}_m^*$ & 90.18 & 0.3240 & 0.0167 & 0.9966 \\
    ${\bf{u}}_m^*$ & \textbf{90.19} & \textbf{0.3091} & \textbf{0.0166} & \textbf{0.9968}\\
    \bottomrule
    \end{tabular}%
  \label{tab_umax}%
\end{table}%

\subsection{Detailed numerical analyses}
We present comprehensive numerical analyses of computational efficiency, segmentation performance, and uncertainty estimation reliability. These can be found in the Tab.~\ref{Tab_time}. Compared to existing uncertainty estimation methods, the integrated DEviS models manifest diminished computational efficiency while showcasing enhanced segmentation robustness. Furthermore, superior calibration outcomes are evident through metrics like ECE and UEO, signifying advancements in reliable medical image segmentation and uncertainty estimation. Specifically, in the 2D skin lesion segmentation task, models such as UE, DU, TTA, and BQNAT, which necessitate multiple predictions, exhibit slightly higher computation times, while the PU method demonstrates substantially less. The U-Net integrated with DEviS demonstrates significantly improved computation time, approximately 198, 88, and 2 times faster than the top three methods UE, BQNAT, and PU, respectively. From Tab.~\ref{Tab_time}., it can be inferred that in the multi-modal 3D brain tumor segmentation task, as input data size and model complexity increase, the computation time of UE escalates, whereas TTA, BQNAT, and PU require relatively less time. In comparison to the U-Net integrated with DEviS, TTA, BQNAT, and PU methods necessitate approximately 687, 399, and 9 times more computation time, respectively. Notably, the nnFormer integrated with DEviS exhibits the highest computational efficiency, with computation times approximately 1060, 616, and 13 times faster than TTA, BQNAT, and PU methods, respectively.

Furthermore, concerning segmentation performance and uncertainty reliability, this paper evaluates segmentation efficacy using the Dice and ASSD metrics. As uncertainty lacks ground truth, the same ECE and UEO metrics as in the literature are employed to indirectly assess the reliability of uncertainty estimation. As delineated in Tab.~\ref{Tab_time}., under noisy conditions, U-Net, Attention-UNet, V-Net, and nnFormer integrated with DEviS demonstrate heightened accuracy and reliability in comparison to existing uncertainty estimation methods.

\begin{table*}
    \centering
    \renewcommand{\arraystretch}{1.1}
    \setlength\tabcolsep{5pt}
    \caption{Inference analysis of Uncertainty estimation models on above two datasets under normal condition and Gaussian noise condition (${\sigma}{\rm{ = }}0.3,0.2$). N and OOD denote the normal condition and out-of-distribution condition, respectively.  \label{Tab_time}
    }\label{tab:Generalizability}
    \begin{tabular}{cr||cccccccccccc}
    \hline
    \rowcolor{mygray}
    & {\multirow{2}{*}{Methods}}  & {\multirow{2}{*}{Testing time$\downarrow$}} & {\multirow{2}{*}{Parameter$\downarrow$}} & {\multirow{2}{*}{FLOPs$\downarrow$}} &  \multicolumn{2}{c}{Dice$\uparrow$}& \multicolumn{2}{c}{ASSD$\downarrow$}& \multicolumn{2}{c}{ECE$\downarrow$} & \multicolumn{2}{c}{UEO$\uparrow$}\\
    & & & & &N & OOD & N & OOD & N & OOD & N & OOD\\
    \hline
    \hline
    \multirow{7}{*}{\begin{sideways}ISIC2018 \end{sideways}} & 
    UE~\cite{ensemble17}  & 13.88 s & \textbf{7.77 M}& 137.34 G & 0.839 & 0.425 & 8.28 & 30.20 & 0.049& 0.142& 0.858&0.416 \\
    &DU~\cite{17dropoutCV} & 26.75 s & \textbf{7.77 M} & 137.34 G & 0.849&0.369 & 8.06&31.78 & 0.052&0.163& 0.865&0.544 \\
    &PU~\cite{2018probabilisticU} & 0.14 s & 27.35 M & 108.54 G & 0.857&0.580 & 8.19&21.87 & 0.055&0.130 & 0.868&0.665\\
    &TTA~\cite{wang2018TTA} & 26.16 s & \textbf{7.77 M} & 137.34 G & 0.870&0.473 & 7.16& 27.02 & 0.048&0.159 & 0.883&0.506\\
    &BQNAT~\cite{roy2019bqnat} & 6.18 s & \textbf{7.77 M} & 41.22 G & 0.851& 0.516 & 7.83 & 32.47 & 0.052 & 0.160 & 0.864 & 0.543\\
    &\textbf{U+Our} & \textbf{0.07 s}& \textbf{7.77 M} & \textbf{13.74 G} & 0.868&0.607 & 7.41& 21.47 & 0.048&0.119 & 0.871&0.674\\
    &\textbf{V+Our} & \underline{0.08 s}& \underline{13.07 M} & 15.45 G & \textbf{0.874}&\underline{0.781}& \textbf{6.71}&\textbf{11.35}& \textbf{0.045}&\textbf{0.087}&\underline{0.905}&\underline{0.839}\\
    &\textbf{nnU+Our} & \underline{0.08 s}& \textbf{7.77 M} & \underline{14.13 G} & \underline{0.873} &\textbf{0.789}& \underline{6.72} & \underline{13.47} & \underline{0.047} &\underline{0.090}&\textbf{0.910}&\textbf{0.853}\\
    \hline
    \hline
    \multirow{10}{*}{\begin{sideways}BraTS2019\end{sideways}} &
    UE \cite{ensemble17}  & 1626.80 s & \underline{4.76 M} & 6941.65 G& 0.857&0.709 & 3.42& 4.93 & 0.0091&0.0211 & 0.857&0.727\\
    &DU \cite{17dropoutCV} & 1428.30 s & \underline{4.76 M} & 6941.65 G& 0.825&0.662 & 2.50 & 5.79  & 0.0062 &0.0104 & 0.855&0.697\\
    &PU \cite{2018probabilisticU} & 15.40 s & 5.13 M & 1423.74 G & 0.864&0.470 & 1.73&10.02 & 0.0058&0.0143 & 0.874&0.622\\
    &TTA~\cite{wang2018TTA} & 1230.09 s & \underline{4.76 M} & 6941.65 G & 0.841&0.682 & 2.54&5.58 & 0.0064&0.0112 & 0.848&0.717\\
    &BQNAT~\cite{roy2019bqnat} & 714.72 s & \underline{4.76 M} & 2082.50 G & 0.846&0.681 & 2.57&4.93 & 0.0062&0.0100 & 0.859&0.701\\
    &U+TBraTS \cite{zou2022tbrats} & \underline{1.68 s}& 4.76 M & 1263.69 G& 0.848&0.698 & 2.62& 5.49 & 0.0062 & 0.0097 & 0.864 & 0.805\\
    &\textbf{U+Our} & 1.79 s& \underline{4.76 M} & 1263.69 G& 0.850&0.718 & 2.46& 4.74 & 0.0058& \underline{0.0088} & 0.866& \underline{0.818}\\
    &\textbf{nnU+Our} & 1.91 s & 16.55 M & 4271.30 G & 0.855 &0.636 & 2.26 & 11.48 & 0.0058 & 0.0185 & 0.867 & 0.636\\
    &\textbf{nnFormer+Our} & \textbf{1.16 s} & 37.61M & \underline{953.99 G} & \textbf{0.873}&0.712 & \underline{1.68} & 4.27 & 0.0056&0.0094 & \underline{0.882} & 0.816\\
    &\textbf{AU+Our} & 1.93 s& 4.77 M & 1285.60 G& 0.859&\textbf{0.803} & 1.89&\textbf{2.25} & \underline{0.0054}&\textbf{0.0071} & 0.880&\textbf{0.845}\\
    &\textbf{V+Our} & 1.71 s & \textbf{2.31 M} & \textbf{790.16 G} & \underline{0.870} &\underline{0.721} & \textbf{1.58}& \underline{4.10} & \textbf{0.0048}&0.0127 & \textbf{0.895}&0.761\\
    \hline
    \end{tabular}
    \end{table*}

\subsection{DEviS Uncertainty in Challenging Case Analysis}
\begin{figure}
\centering
\includegraphics[width=1\linewidth]{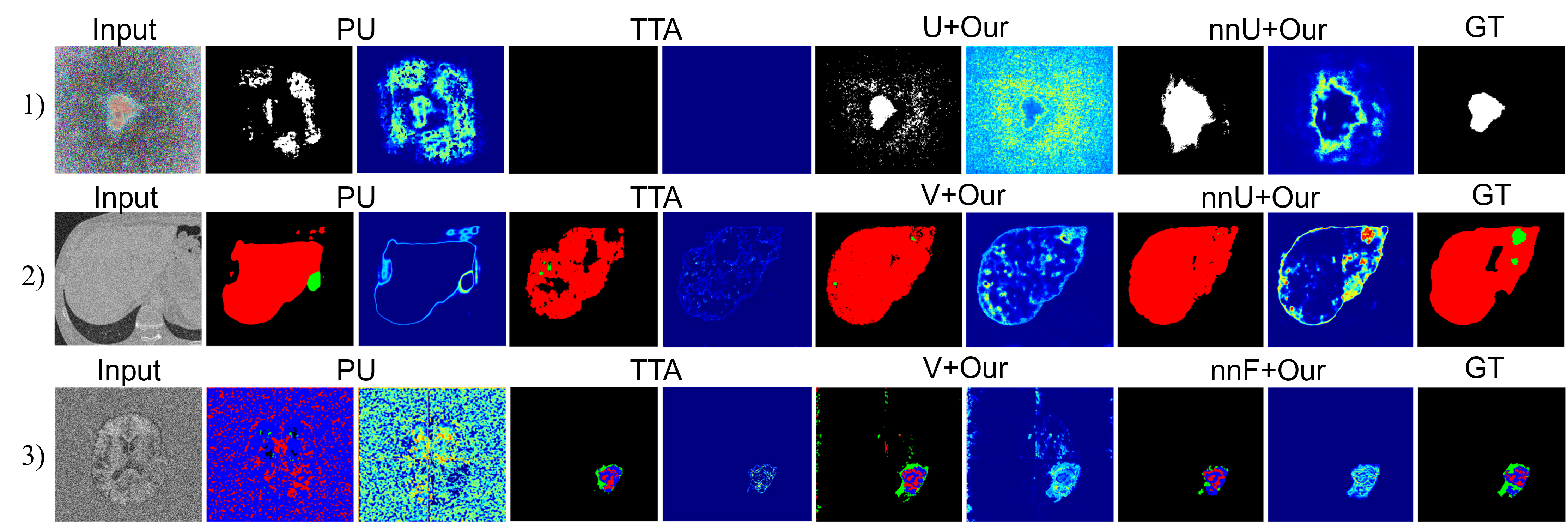}
\caption{Uncertainty analysis on three representative failure cases from ISIC2019, LiTS2017, and BraTS2018.}
\label{F_Challenging}
\end{figure}

We further analyze several failure cases of uncertainty estimation in Figure~\ref{F_Challenging}. Each case in the figure corresponds to a test image with added high-level noise from the three datasets described in this study, namely ISIC2018, LiTS2017, and BraTS2019. In Fig.~\ref{F_Challenging} 1), although nnU-Net with our method highlights some incorrect boundaries, false positives remain, and uncertainty is low in central regions due to surrounding noise. Compared to PU and TTA, U-Net shows more robust performance in both prediction and uncertainty. In Fig.~\ref{F_Challenging} 2), V-Net with our method fails to detect a small tumor in the lower-right GT region, with low uncertainty at the missed location. In contrast, nnU-Net better captures this uncertainty. PU and TTA also show limited sensitivity in this case. In Fig.~\ref{F_Challenging} 3), V-Net produces false positives with low uncertainty, and inter-class uncertainty remains insufficient. Despite these issues, our method shows improved prediction quality and uncertainty estimation compared to PU and TTA. Overall, while challenges remain in handling false positives, small targets, and ambiguous regions, our method provides more reliable uncertainty estimation than existing baselines.

\subsection{Intuitive segmentation results on the clinical application}
\begin{figure}
\centering
\includegraphics[width=1\linewidth]{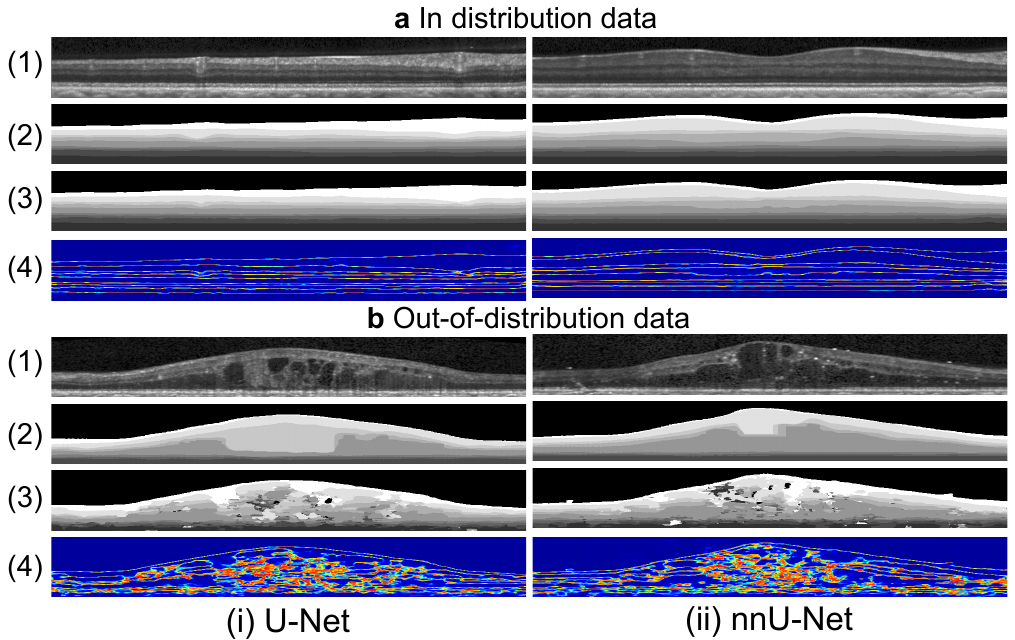}
\includegraphics[width=1\linewidth]{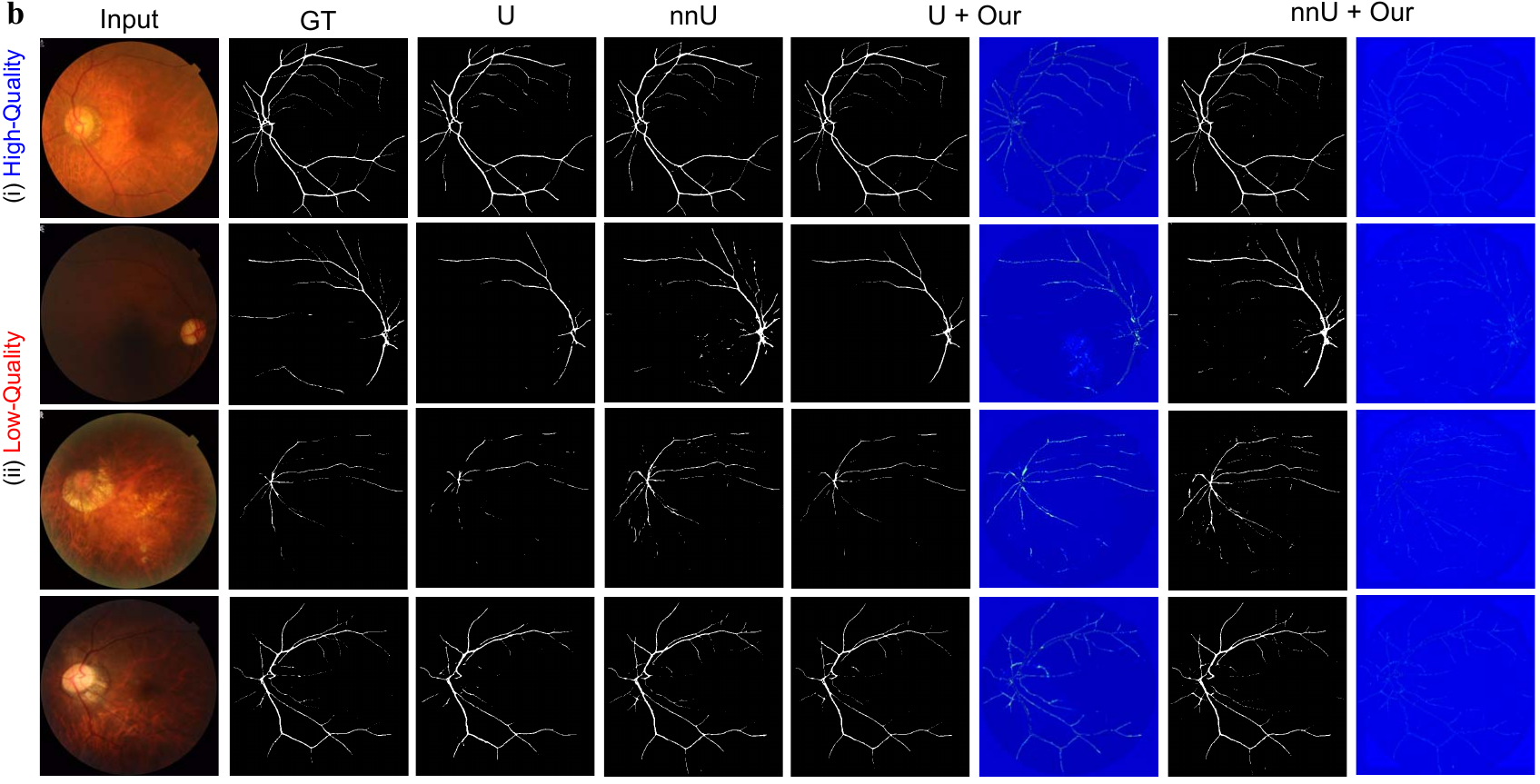}
\caption{Visualization results of clinical applications. \textbf{a.} ID (JH-OCT dataset without disease) and \textbf{b.} OOD data (Duke-OCT dataset with DME). (1) Input image (2) Ground-truth (3-4) Prediction and uncertainty map of U-Net and nnU-Net with Our DEviS. \textbf{b.} High and Low-quality data visualization results of baselines (U-Net and nnU-Net) with our DEviS on the FIVES dataset. }
\label{A_F5}
\end{figure}

For a more intuitive illustration, segmentation results and uncertainty maps for both ID and OOD data are shown in Fig.\ref{A_F5}. These results indicate that DEviS with UAF can effectively identify and filter abnormal regions where lesions may be present in OOD data. Additionally, Fig.\ref{A_F5}\textbf{b} presents segmentation results and uncertainty maps for low- versus high-quality images, providing a clearer visualization of quality differences. Collectively, these findings demonstrate that DEviS with UAF can serve as a reliable image quality indicator, facilitating fair valuation of personal data in healthcare and consumer applications, by helping to remove low-value or harmful data while prioritizing high-value data for diagnostic support.

\end{document}